\documentclass[aps,pra,twocolumn,superscriptaddress,floatfix]{revtex4-1}

\usepackage{graphicx}
\usepackage{dcolumn}
\usepackage{bm}
\usepackage{amssymb}
\usepackage{microtype}
\usepackage{xfrac}
\usepackage{makecell}
\usepackage{array}
\usepackage[version=4]{mhchem}
\usepackage{gensymb}
\usepackage{multirow}
\usepackage[colorlinks=true]{hyperref}
\newcommand{\s}{\mathbf S}

\begin{document}

\title{Theory of spin waves in a hexagonal antiferromagnet}

\author{S.~Dasgupta}
\affiliation{Institute for Quantum Matter and Department of Physics and Astronomy, Johns Hopkins University, Baltimore, MD 21218, USA}

\author{O.~Tchernyshyov}
\affiliation{Institute for Quantum Matter and Department of Physics and Astronomy, Johns Hopkins University, Baltimore, MD 21218, USA}

\begin{abstract} 
We construct a field-theoretic description of spin waves in hexagonal antiferromagnets with three magnetic sublattices and coplanar $120^\circ$ magnetic order. The three Goldstone modes can be separated by point-group symmetry into a singlet $\alpha_{0}$ and a doublet $\bm{\beta}$. The $\alpha_0$ singlet is described by the standard theory of a free relativistic scalar field. The field theory of the $\bm{\beta}$ doublet is analogous to the theory of elasticity of a two-dimensional isotropic solid with distinct longitudinal and transverse ``speeds of sound.'' The well-known Heisenberg models on the triangular and kagome lattices with nearest-neighbor exchange turn out to be special cases with accidental degeneracy of the spin-wave velocities. The speeds of sound can be readily calculated for any lattice model. We apply this approach to the compounds of the Mn$_3$X family with stacked kagome layers. 
\end{abstract}

\maketitle

\section{Introduction}\label{sec.Intro}
The study of spin waves, gentle excitations around a magnetic ground state, in terms of a local, continuum field theory is well established \cite{Aharoni1996}. The ordered moments are expressed in terms of a classical spin, or magnetization, field $\mathbf m(t,\mathbf r)$. Although this approach cannot be applied on the atomic scale, it has proved to be useful to study the slow spatial and temporal fluctuations of magnetization. These field theories have been extensively studied for simple magnets with one or two magnetic sublattices \cite{haldane-o3-1dafm,haldane-o3-2dafm}. In these highly symmetric scenarios the emergent field theory is a non-linear $\sigma$ model for an appropriate order parameter. The field theory has also been utilized to study the combined interactions of spin waves (magnons) with solitons like domain walls \cite{skkim14}, and magnetic vortices \cite{sheka18}.

The spin waves are conveniently expressed in the basis of normal modes of the spin system. These modes form a symmetry governed irreducible representation for the spin degrees of freedom (rotational) in a magnetic unit cell \cite{georgi99}. The normal modes in the case of a system where exchange is the dominant interaction, provide an intuitive picture of the spin wave excitations. In addition they provide insight into how the spin order couples to internal anisotropies and external perturbations, based on symmetry arguments.

In antiferromagnets the exchange interaction enforces a zero net magnetization per unit cell, $\sum_{i} \mathbf{S}_i = 0$, where the summation is over sublattices. Normal modes that violate this condition are costly and will be referred to as `hard.' We will focus on soft modes that preserve the condition of zero net spin, they enter the energy density $\mathcal{U}$ in the form of gradients. 

In this paper we construct the spin-wave theory for generic hexagonal antiferromagnets with three magnetic sublattices. Previous field-theoretic treatments include the works of \textcite{dombre1988,Dombre1989} and \textcite{Mineev1996}. We adopt the continuum approach to study the universal features of this class of magnets associated with its soft modes, long-wavelength spin waves. The triangular-lattice \cite{Chubukov1999} and kagome \cite{Harris1992} antiferromagnets with exchange between nearest neighbors only turn out to be special cases with accidental degeneracy of the spin-wave spectra. 

Some features unique to the three-sublattice antiferromagnet emerge from this construction. Firstly, there are now three Goldstone modes as compared to two for the two-sublattice case. This happens because the N\'{e}el order parameter (staggered magnetization) for the two-sublattice case breaks the $SO(3)$ symmetry of the spin vectors only partially, down to $SO(2)$ rotations about the N\'{e}el vector. The three-sublattice magnetic order breaks the symmetry fully, resulting in three Goldstone modes. Secondly, from the perspective of point-group symmetry, the three Goldstone modes can be partitioned into a singlet and a doublet. The field theory for this doublet turns out to be analogous to the continuum theory of elasticity in two dimensions.

We start by reviewing the familiar micromagnetic field theories of the easy-plane Heisenberg ferromagnet \cite{Kosterlitz1974,dasgupta2019quantum} and the two-sublattice Heisenberg antiferromagnet in Sec.~\ref{sec.one-two-sl}. We use these familiar settings to illustrate some important but rarely discussed issues such as the emergence of kinetic energy for soft modes coupled through the Berry phase to hard modes. We proceed to a study of the lattice geometry and normal mode structure in hexagonal antiferromagnets, Sec.~\ref{sec.three-sl}. We derive a field theory for the soft modes and test this theory on the familiar kagome and triangular-lattice antiferromagnets in Sec.~\ref{sec.ped_examples}, comparing our results with the Holstein Primakoff calculations on these models \cite{Dombre1989,Chubukov1999,Harris1992}. In Sec.~\ref{sec.Stacked_kagome}, we apply our approach to the stacked kagome antiferromagnets of the Mn$_3$X family (where X = Ge, Sn). Although structurally complex, the basic magnetic unit of this system is the triangular antiferromagnet forming the basis for our theory to be applied to obtain its spin wave spectrum. We discuss the broader applicability of our effective field theory in Sec.\ref{sec.conclusion}. Some of the more technical parts are collected in the Appendix.

\section{One and two sublattice field theories in 2d}\label{sec.one-two-sl}
\subsection{Easy-plane ferromagnet}
The lattice model of an easy-plane ferromagnet with nearest-neighbor Heisenberg exchange $J>0$ and local anisotropy $K$ has the potential energy
\begin{eqnarray}
\label{eq.fm-energy-density}
U = 
- J\sum_{<ij>}\mathbf{S}_i\cdot\mathbf{S}_j + \frac{K}{2} \sum_{i}S_{iz}^{2}.
\end{eqnarray}
Here $i$ and $j$ denote lattice sites and $\langle ij \rangle$ a nearest-neighbor bond. At distances much greater than the lattice constant $a$, we may use a continuum theory where discrete spins $\mathbf S_i$ of length $S$ are represented by a spin vector field $\mathbf m(\mathbf r)$ of unit length smoothly varying in space:
\begin{equation}
\mathbf S_i  
\approx S \, \mathbf m(\mathbf r_i), \end{equation} 
where $\mathbf r_i$ is the position of spin $\mathbf S_i$ in the lattice. The length constraint $|\mathbf m| = 1$ can be resolved by expressing the spin field $\mathbf m$ in terms of the polar and azimuthal angles $\theta$ and $\phi$, 
\begin{equation}
\mathbf m = (\sin{\theta} \cos{\phi}, 
    \sin{\theta} \sin{\phi},
    \cos{\theta}).    
\end{equation}

The energy (\ref{eq.fm-energy-density}) is dominated by Heisenberg exchange, which it is convenient to rewrite as follows: 
\begin{equation}
U_\text{ex} 
= - J \sum_{\langle ij\rangle}\mathbf S_i \cdot \mathbf S_j 
= J\sum_{\langle ij\rangle}
\frac{1}{2}(\mathbf S_i - \mathbf S_j)^2 - S^2.    
\end{equation}
It is evidently minimized by a uniform state with all spins $\mathbf S_i$ 
pointing in the same direction. In the continuum limit, the spin difference in the exchange energy is approximated by a spatial gradient,
\begin{equation}
\mathbf S_i - \mathbf S_j 
\approx (\mathbf r_i - \mathbf r_j) \cdot \nabla S \mathbf m.
\end{equation}
The energy functional $U[\mathbf m(\mathbf r)] =  \int d^2r \, \mathcal U$ has the energy density 
\begin{eqnarray}
\mathcal U 
&=& \frac{\mathcal{J}}{2}(\nabla\mathbf{m})^{2} 
    + \frac{\mathcal K }{2} m_{z}^{2} 
\nonumber\\
&=& \frac{\mathcal{J}}{2}
    \left[
        (\nabla\theta)^{2}
        + \sin^2{\theta} (\nabla\phi)^{2}
    \right]
    + \frac{\mathcal K}{2} \cos^2{\theta}.
\label{eq:U-easy-plane-ferromagnet-full}
\end{eqnarray}
Here $(\nabla\mathbf{m})^{2} \equiv \sum_n \partial_n \mathbf m \cdot \partial_n \mathbf m$ with a summation over the spatial Cartesian indices $n=x,y$. The coupling constants of the continuum theory are related to those of the lattice model. For a square lattice, $\mathcal J = JS^2$ and $\mathcal K = KS^2/a^2$. Another important quantity is the density of angular momentum (spin) $\mathcal S = S/a^2$ on a square lattice. 

The dynamics of the spin field is governed by the Landau-Lifshitz equation, equating the rate of change of the angular momentum to the local torque:
\begin{equation}
\label{eq:LLG}
\mathcal S \, \dot{\mathbf m} = 
	- \mathbf m \times 
	\frac{\delta U}{\delta \mathbf m}.
\end{equation}
Here $\delta U[\mathbf m(\mathbf r)]/\delta \mathbf m(\mathbf r)$ is the functional derivative of the energy.

The Landau-Lifshitz equation (\ref{eq:LLG}) can be derived from a Lagrangian $\mathcal{L} =  \mathcal L_B - \mathcal{U}$, where 
\begin{equation}
\label{eq.L-B}
\mathcal L_B 
\equiv \mathcal S \, \mathbf a(\mathbf m) \cdot\dot{\mathbf m}
\end{equation}
is a kinematic term originating from the spin Berry phase  \cite{shankar2017,dasgupta18}. The vector potential $\mathbf{a} (\mathbf{m})$ represents the magnetic field of a monopole on the spin unit sphere,  $\nabla_{\mathbf{m}}\times\mathbf{a} = - \mathbf{m}$. The standard choices for the vector potential, 
\begin{equation}
\mathcal L_B = \mathcal S (\cos{\theta} \pm 1)\dot{\phi}    
\label{eq.L-B-standard}
\end{equation}
have a Dirac-string singularity at the north and south pole, respectively.

The full Lagrangian of the continuum theory, 
\begin{equation}
\mathcal L = 
\mathcal S (\cos{\theta} \pm 1)\dot{\phi}
- \frac{\mathcal{J}}{2}
    \left[
        (\nabla\theta)^{2}
        + \sin^2{\theta} (\nabla\phi)^{2}
    \right]
- \frac{\mathcal K}{2} \cos^2{\theta},
\label{eq:L-easy-plane-ferromagnet-full}
\end{equation}
yields characteristic length and time scales of the model,
\begin{equation}
\ell_0 = \sqrt{\mathcal J/\mathcal K},
\quad
t_0 = \mathcal S/\mathcal K.
\label{eq:length-time-scales}
\end{equation}
The anisotropy is usually much weaker than exchange, so that $\ell_0 \gg a$. 

The easy-plane anisotropy forces the spins to stay close to the equatorial plane, $\theta \approx \pi/2$, making the polar angle $\theta$ a hard mode. However, setting $\theta = \pi/2$ is not a good idea because doing so would rob the soft mode $\phi$ of its dynamics. [The surviving kinetic term $\pm S\dot{\phi}$ in Eq.~(\ref{eq:L-easy-plane-ferromagnet-full}) would not contribute to the classical equation of motion for $\phi$.]

Instead of neglecting the hard mode $\theta$ altogether, we proceed to eliminate it more carefully. As long as we are interested in the slow dynamics of the system (on length and time scales longer than $\ell_0$ and $t_0$), we may neglect the gradient term $(\nabla \theta)^2$ and set $\sin^2{\theta} = 1$ in Eq.~(\ref{eq:L-easy-plane-ferromagnet-full}). In this slow limit, the Lagrangian simplifies to 
\begin{equation}
\mathcal L = 
\mathcal S (\cos{\theta} \pm 1)\dot{\phi}
- \frac{\mathcal{J}}{2} (\nabla\phi)^{2}
- \frac{\mathcal K}{2} \cos^2{\theta}.
\label{eq:L-easy-plane-ferromagnet-full-slow}
\end{equation}
From it, we obtain the equation of motion for the hard field $\theta$: 
\begin{equation}
\mathcal S \dot{\phi} - \mathcal K \cos{\theta} = 0. 
\end{equation} 
It can be seen that, in the slow limit, the hard mode $\theta$ instantaneously adapts to the velocity of the soft mode $\dot{\phi}$. This allows us to eliminate the hard field $\theta$ and obtain the following Lagrangian for the soft field $\phi$:
\begin{equation}
\mathcal L(\phi) =  \frac{\rho}{2} \dot{\phi}^2 
    - \frac{\mathcal{J}}{2} (\nabla \phi)^2,
\label{eq:L-low-energy}
\end{equation}
where $\rho = \mathcal S^2/\mathcal K$. The slow dynamics of the $\phi$ field is described by the wave equation $\rho \ddot{\phi} = \mathcal J \nabla^2 \phi$ with the characteristic velocity $v = \sqrt{\mathcal J/\rho} = \ell_0/t_0$. 

Our procedure of integrating out the hard mode $\theta$ produced a kinetic energy in the effective Lagrangian (\ref{eq:L-low-energy}) of the soft field $\phi$. This emergent inertia is common in ferromagnets and is known as the D\"{o}ring mass \cite{doring1948}. The elimination of the hard mode $\theta$ is justified on length and time scales longer than the characteristic ones (\ref{eq:length-time-scales}). For fast processes, we have to retain the hard mode $\theta$ and the full Lagrangian (\ref{eq:L-easy-plane-ferromagnet-full}).

\subsection{Two-sublattice antiferromagnet}
Our next familiar example is the Heisenberg antiferromagnet on the square lattice with the energy
\begin{equation}
\label{eq.Heisenberg-afm}
U = J \sum_{<ij>}\mathbf{S}_i\cdot\mathbf{S}_j,
\end{equation}
where $J>0$ is the strength of nearest-neighbor antiferromagnetic exchange. In the ground state, spins of the two sublattices point in the opposite directions. For this reason, we must use two slowly varying spin vector fields of unit length $\mathbf m_1(\mathbf r)$ and $\mathbf m_2(\mathbf r)$ for a continuum description, one for each sublattice. Proceeding along the same lines as with the ferromagnet, we obtain the following energy density in the continuum approximation:
\begin{equation}
\mathcal U = 
JS^2
\left(
    \frac{2\mathbf m_1 \cdot \mathbf m_2}{a^2}
    - \frac{1}{2}\nabla \mathbf m_1 \cdot \nabla \mathbf m_2
\right).
\end{equation}

Because $\mathbf m_1 = - \mathbf m_2$ in a ground state, it is tempting to approximate $\mathbf m_2 \approx - \mathbf m_1$ at low energies, which would reduce the number of independent variables. We will eventually accomplish that. However, the process requires some care. We proceed gradually and at first introduce two new fields, the uniform and staggered spin: 
\begin{equation}
\mathbf m = \mathbf m_1 + \mathbf m_2,
\quad
\mathbf n = \frac{\mathbf m_1 - \mathbf m_2}{2}.
\end{equation}
At low energies, the uniform spin field is suppressed, $\mathbf m \approx 0$, so this field represents a hard degree of freedom and we will eventually integrate it out. The staggered field $\mathbf n$ will represent the spin fields of both sublattices, 
\begin{equation}
\mathbf m_1 = \frac{\mathbf m}{2} + \mathbf n \approx \mathbf n,
\quad
\mathbf m_2 = \frac{\mathbf m}{2} - \mathbf n \approx -\mathbf n.
\end{equation}

The length constraints $|\mathbf m_1|^2 = |\mathbf m_1|^2 = 1$ translate into the following constraints on the new fields:
\begin{equation}
\mathbf m \cdot \mathbf n = 0,
\quad
\mathbf n^2  
+ \frac{\mathbf m^2}{4} = 1.
\end{equation}

The energy density, measured relative to the ground state and expressed in terms of the uniform and staggered spin fields, reads
\begin{equation}
\mathcal U = 
\frac{JS^2}{2}    
\left(
    \frac{2\mathbf m^2}{a^2}
    - \frac{1}{4} (\nabla \mathbf m)^2
    + (\nabla \mathbf n)^2
\right).
\end{equation}
The first term expresses the main effect of antiferromagnetic exchange: it suppresses the uniform spin $\mathbf m$. The second term is comparatively small for slow spatial variations of $\mathbf m$ and may therefore be neglected. In contrast, the staggered field enters the energy through gradient terms only.

These considerations motivate the following simplified form of the energy density: 
\begin{equation}
\mathcal U = 
\frac{\mathbf m^2}{2 \chi} 
+ \frac{\mathcal J}{2} (\nabla \mathbf n)^2.
\end{equation}
Here $\chi = a^2/2JS^2$ is, up to a multiplicative constant, magnetic susceptibility and $\mathcal J = JS^2$ is the continuum exchange constant.

The kinetic term in the Lagrangian originates from the Berry phases of spins from both sublattices,
\begin{equation}
\label{eq.kin-L-1}
\mathcal{L}_{B} = 
\mathcal S 
\left[
    \mathbf{a}_{1}(\mathbf m_1) 
        \cdot \dot{\mathbf{m}}_{1} 
    + \mathbf{a}_{2}(\mathbf m_2) 
        \cdot \dot{\mathbf{m}}_{2}
\right], 
\end{equation}
where $\mathcal{S} = S/2a^2$ is the spin density on one sublattice.

A judicious choice of the gauge potentials $\mathbf a_1$ and $\mathbf a_2$ yields the following simple result \cite{ivanov1995, kim2014}: 
\begin{equation}
\mathcal L_B = 
\mathcal{S} \, \mathbf m
\cdot (\dot{\mathbf n} \times \mathbf n).    
\end{equation}
We see from it why setting $\mathbf m = 0$ at the outset would be a bad idea: we would lose the kinetic term of the Lagrangian. Like the polar angle $\theta$ in the easy-plane ferromagnet, the uniform spin $\mathbf m$ is a hard mode. However, these hard modes mediate the dynamics of the soft modes $\phi$ and $\mathbf n$, respectively. 

We thus arrive at the effective Lagrangian of the antiferromagnet for the uniform and staggered spin fields,
\begin{equation}
\mathcal{L} =
\mathcal{S} \, \mathbf m
\cdot (\dot{\mathbf n} \times \mathbf n)  
- \frac{\mathbf m^2}{2\chi} 
- \frac{\mathcal J}{2} (\nabla \mathbf n)^2.
\end{equation}

As in the previous example, we use the equation of motion for the hard mode $\mathbf m$,
\begin{equation}
\mathbf m = 
\chi \mathcal S \, \dot{\mathbf n} \times \mathbf n,
\end{equation}
to eliminate it and to thereby obtain an effective kinetic energy for the soft mode $\mathbf n$: 
\begin{equation}
    \mathcal{L}_\text{kin} = \frac{\rho}{2}(\dot{\mathbf{n}}\times\mathbf{n})^2 = \frac{\rho}{2}\dot{\mathbf{n}}^2.
\end{equation}
Here we have used the orthogonality of the unit vector $\mathbf n$ and its derivative $\dot{\mathbf n}$ to simplify the expression. $\rho = \mathcal S^2 \chi$ is the inertia density for the staggered spin $\mathbf n$.  

The full Lagrangian of the soft mode $\mathbf n$ is
\begin{equation}
    \mathcal{L} = \frac{\rho}{2}\dot{\mathbf n}^{2} - \frac{\mathcal J}{2}\left(\nabla\mathbf n\right)^2,
\end{equation}
There are two degenerate Goldstone modes, representing oscillations of $\mathbf n$ in the two directions orthogonal to its ground-state orientation.  Both modes disperse linearly according to $\omega = v k$, with the speed $v = \sqrt{\mathcal J/\rho} = 2\sqrt{2} J S a$ on the square lattice.

\section{Three-sublattice Field Theory in 2d}
\label{sec.three-sl}
\begin{figure}[t]
    \includegraphics[width=\columnwidth]{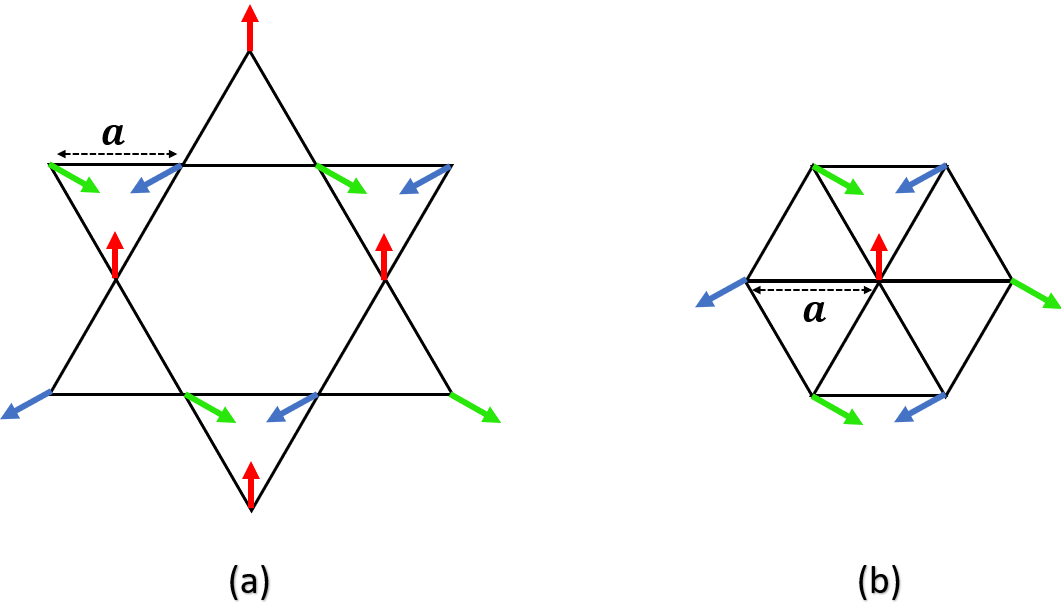}
    \centering
    \caption{The two prototype triangular lattices and their lattice parameters, the kagome lattice of corner sharing triangles (a) and the triangular lattice (b). We show a 120$^{\circ}$ ordered state on both where sites with the same spin color or orientation belong to the same spin sublattice.}
    \label{fig:lattice}
\end{figure}

Antiferromagnets with a hexagonal or trigonal lattice symmetry often exhibit strong geometrical frustration, manifested in their inability to form a N{\'e}el magnetic order with just two sublattices. Well-known examples are the Heisenberg antiferromagnet on a triangular lattice and on kagome (Fig.~\ref{fig:lattice}), whose ground states have three magnetic sublattices. Magnets of this class share robust common features such as the existence of three Goldstone modes, spin waves with a linear dispersion $\omega \sim v k$ in the long-wavelength limit $k \to 0$. Their existence is related to the spontaneous breaking of the SO(3) spin-rotation symmetry.

Anisotropic interactions, induced by spin-orbit coupling and dipolar interactions, explicitly break the SO(3) symmetry and open gaps in the spin-wave spectra. However, these anisotropies are typically weak in comparison with Heisenberg exchange. Therefore, this symmetry exists in at least an approximate form and the picture of three Goldstone modes with a linear dispersion is a good starting point.

\subsection{Lattice and spin geometry}
\begin{figure}[t]
\includegraphics[width=\columnwidth]{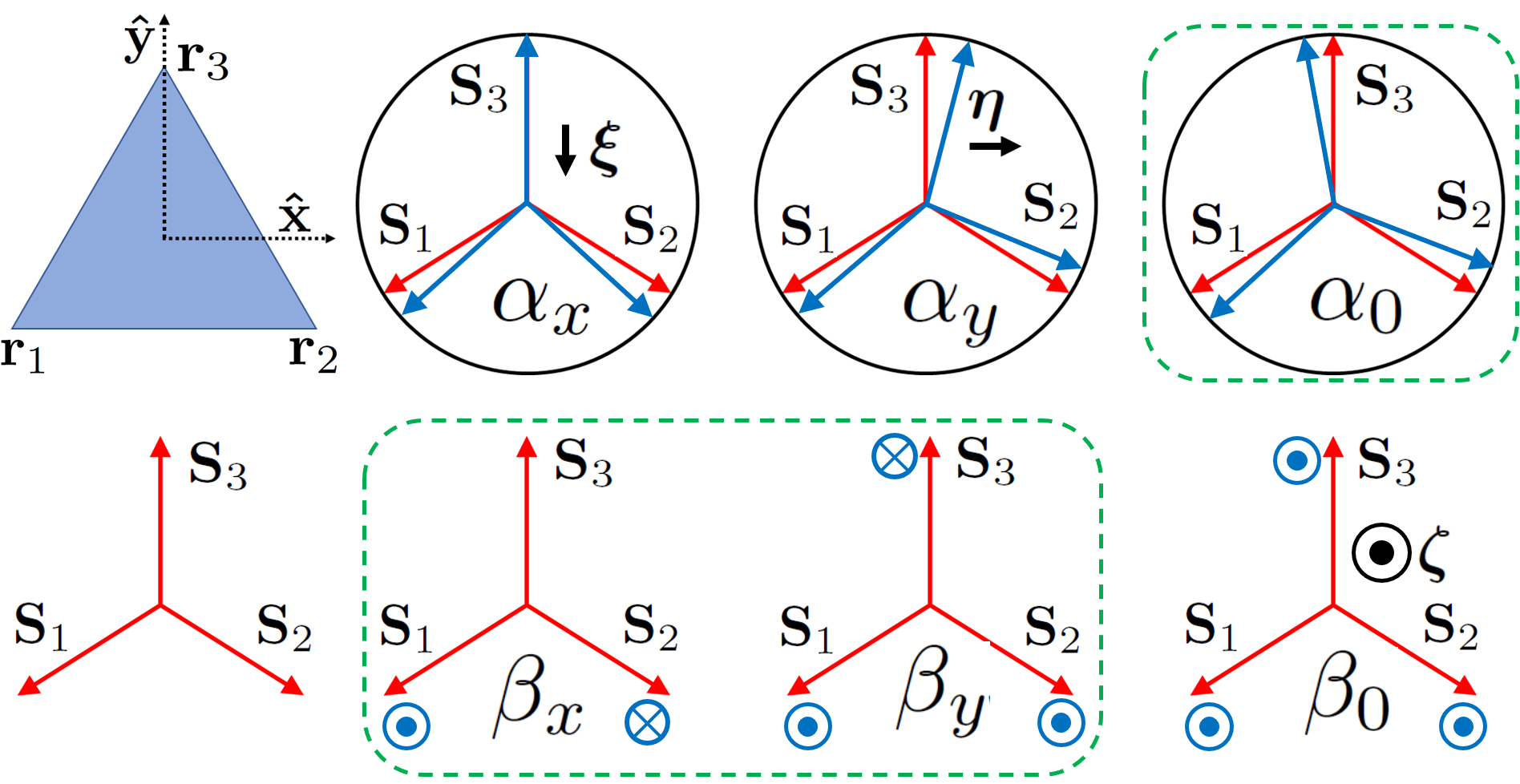}
\centering
\caption{ On the left we have the geometry of a single triangular plaquette with an example 120$\degree$ ground state. The spins carry the same labels as the site i.e spin $\mathbf{S}_i$ is at site $\mathbf{r}_i$.  The normal modes for the spin structure are on the right. The red arrow indicates the ground state, while the blue arrows indicate the distorted state. The green dashed boxes mark the soft modes, $(\alpha_0,\bm{\beta})$.}
    \label{fig:lattice_modes}
\end{figure}

The magnetic unit cell has three sites forming an equilateral triangle (Fig.~\ref{fig:lattice_modes}). In our convention, the sublattice index $i=1, 2, 3$ increases as we go around the unit cell counterclockwise.

The spins $\mathbf S_1$, $\mathbf S_2$, and $\mathbf S_3$ interact with one another via antiferromagnetic Heisenberg exchange of equal strength so that the net spin of the cell vanishes in the ground state,
\begin{equation}
\mathbf S_1 + \mathbf S_2 + \mathbf S_3 = 0.
\label{eq:vacuum}
\end{equation}
In the exchange approximation, spin and lattice rotations are decoupled and we may consider spatial symmetries separately. The point group of the magnetic unit cell is the dihedral group $D_3$, the group of the equilateral triangle. Spatial rotations through the angle $+2\pi/3$ about the $c$ axis produce a cyclic exchange of the spin variables: 
\begin{equation}
\left(
    \begin{array}{c}
        \s_1'\\
        \s_2'\\
        \s_3'
    \end{array}
\right)    
= 
\left(
    \begin{array}{ccc}
        0 & 0 & 1\\
        1 & 0 & 0\\
        0 & 1 & 0
    \end{array}
\right) 
\left(
    \begin{array}{c}
        \s_1\\
        \s_2\\
        \s_3
    \end{array}
\right). 
\label{eq:120-rotation-spins}
\end{equation}
Note that the spins $\s_i$ are permuted but there is no rotation in spin space. A $\pi$ rotation about the $b$ axis exchanges spins 1 and 2: 
\begin{equation}
\left(
    \begin{array}{c}
        \s_1'\\
        \s_2'\\
        \s_3'
    \end{array}
\right)    
= 
\left(
    \begin{array}{ccc}
        0 & 1 & 0\\
        1 & 0 & 0\\
        0 & 0 & 1
    \end{array}
\right) 
\left(
    \begin{array}{c}
        \s_1\\
        \s_2\\
        \s_3
    \end{array}
\right). 
\label{eq:mirror-reflection-spins}
\end{equation}
We shall make use of the point group later, when we classify the normal modes by its irreducible representations.

The ground-state condition (\ref{eq:vacuum}) indicates that the three spins are coplanar. The spin plane can be arbitrary in a model with exchange interactions only, which respect the global SO(3) spin-rotation symmetry. Weak anisotropic interactions break this symmetry and favor some special planes. The most common easy plane is the $ab$ plane of a hexagonal or trigonal lattice perpendicular to the sixfold or threefold rotation axis $c$ as is the case in Mn$_3$Ge \cite{Chen20}, Mn$_3$Sn \cite{Nakatsuji:2015}. In some cases spins do orient perpendicular to the ab plane as in NaYbO$_2$ \cite{bordelon2020}. We shall assume that $ab$ is the easy plane in what follows.

The spin reference frame is defined by three orthogonal unit vectors $\{\hat{\bm\xi}, \bm \eta, \bm \zeta\}$ chosen as follows. $\hat{\bm\xi}$ is parallel to $-\mathbf S_3$, $\hat{\bm\eta}$ points along $\mathbf S_2 - \mathbf S_1$, and $\hat{\bm\zeta} = \hat{\bm\xi} \times \hat{\bm\eta}$, Fig. \ref{fig:lattice_modes}.

With the spins in the easy $ab$ plane, there remain two degrees of freedom to change their orientations, one discrete and the other continuous. The discrete degree of freedom, vorticity $q = \pm 1$, specifies how the spins rotate as we go around the triangle. The \emph{spatial} rotation taking site 1 to site 2 is by the angle $2\pi/3$ about the $c$ axis. The \emph{spin} rotation taking $\mathbf S_1$ into $\mathbf S_2$ is by the angle $2\pi/3$ about $\hat{\bm\zeta}$, or by $2q\pi/3$ about $\hat{\mathbf c} = q \hat{\bm\zeta}$. The ground states with $q=+1$ and $-1$ and may be called ``vortex'' and ``antivortex'' states, respectively. The remaining continuous degree of freedom is a rotation within the easy plane.

Cartesian spin components in the reference frame are conveniently expressed in terms of the polar and azimuthal angles $\theta$ and $\phi$:
\begin{equation}
S_\xi = S \sin{\theta} \cos{\phi}, 
\ 
S_\eta = S \sin{\theta} \sin{\phi},
\ 
S_\zeta = S \cos{\theta}.
\end{equation}

\subsubsection{Normal modes}

It is convenient to express the three pairs of spherical angles in terms of six normal modes $\alpha_0$, $\alpha_x$, $\alpha_y$, $\beta_0$, $\beta_x$, and $\beta_y$, see Fig.~\ref{fig:lattice_modes}(b):
\begin{eqnarray}
\label{eq.gs.normal.modes}
\left(
    \begin{array}{c}
        \phi_1\\
        \phi_2\\
        \phi_3
    \end{array}
\right)
&=&  q\left(
    \begin{array}{c}\vspace{.2cm}
         \frac{2\pi}{3}\\ \vspace{.2cm}
         \frac{4\pi}{3}\\
              0
    \end{array}
\right)
- qR
\left(
    \begin{array}{c}\vspace{.2cm}
        \alpha_x\\ \vspace{.2cm}
        \alpha_y\\
        \alpha_0 
    \end{array}
\right), \\ \nonumber
\left(
    \begin{array}{c}
        \theta_1\\
        \theta_2\\
        \theta_3
    \end{array}
\right)
&=& \left(
    \begin{array}{c}\vspace{.2cm}
        \frac{\pi}{2}\\\vspace{.2cm}
        \frac{\pi}{2}\\
        \frac{\pi}{2}
    \end{array}
\right)
+R
\left(
    \begin{array}{c}\vspace{.2cm}
        \beta_x \\\vspace{.2cm}
        \beta_y \\
        \beta_0 
    \end{array}
\right),
\end{eqnarray}
where $R$ is the orthogonal matrix.
\begin{equation}
R =
\left(
    \begin{array}{ccc}\vspace{.2cm}
        \frac{1}{\sqrt{2}} & \frac{1}{\sqrt{6}} & \frac{1}{\sqrt{3}}\\\vspace{.2cm}
        -\frac{1}{\sqrt{2}} & \frac{1}{\sqrt{6}} & \frac{1}{\sqrt{3}} \\
        0 & -\frac{2}{\sqrt{6}} & \frac{1}{\sqrt{3}}
    \end{array}
\right).
\end{equation}

Under spatial transformations of the point group $D_3$, the modes $\alpha_0$ and $\beta_0$ stay unchanged. We therefore call them scalar modes. Modes $\alpha_x$ and $\alpha_y$ form a doublet transforming as 2 components of a vector. Under the $+2\pi/3$ rotation (\ref{eq:120-rotation-spins}), 
\begin{equation}
\left(
    \begin{array}{c}
        \alpha_x'\\
        \alpha_y'
    \end{array}
\right)
= 
\left(
    \begin{array}{rr}
        -\frac{1}{2} & -\frac{\sqrt{3}}{2}\\
        \frac{\sqrt{3}}{2} & -\frac{1}{2}
    \end{array}
\right)
\left(
    \begin{array}{c}
        \alpha_x\\
        \alpha_y
    \end{array}
\right).
\end{equation}
Under the $\pi$ rotation (\ref{eq:mirror-reflection-spins}),
\begin{equation}
\left(
    \begin{array}{c}
        \alpha_x'\\
        \alpha_y'
    \end{array}
\right)
= 
\left(
    \begin{array}{rr}
        -1 & 0\\
        0 & 1
    \end{array}
\right)
\left(
    \begin{array}{c}
        \alpha_x\\
        \alpha_y
    \end{array}
\right).
\end{equation}
The same applies to the modes $(\beta_x, \beta_y)$ which also form a doublet. 

Thus we can separate the normal modes into two scalars (singlets) $\alpha_0$ and $\beta_0$ and two vectors (doublets) $\bm{\alpha} = (\alpha_x,\alpha_y)$ and $\bm{\beta} = (\beta_x,\beta_y)$.

\subsubsection{Canonical pairs}
\label{sec:hexagonal-geometry-canonical}

Modes $\alpha_x$, $\alpha_y$, and $\beta_0$ are proportional to the net spin on a triangle in directions $\hat{\bm\xi}$, $\hat{\bm\eta}$, and $\hat{\bm\zeta}$, respectively, Fig.~\ref{fig:lattice_modes}(b). The other three modes, $\beta_x$, $\beta_y$, and $\alpha_0$, quantify spin rotations about directions $-\hat{\bm\xi}$, $-\hat{\bm\eta}$, and $\hat{\bm\zeta}$. The angle of rotation about the $\hat{\bm\zeta}$ axis is $\alpha_0/\sqrt{3} \equiv \phi_0$.

One may think of $-\beta_x$, $-\beta_y$, and $\alpha_0$ as of global rotation angles and of $\alpha_x$, $\alpha_y$, and $\beta_0$ as of the corresponding components of angular momentum, along the lines of \textcite{Mineev1996}. This also means that $\{-\beta_x,\alpha_x\}$, $\{-\beta_y,\alpha_y\}$, and $\{\alpha_0,\beta_0\}$ are canonical pairs.

\subsubsection{Hard and soft modes}

By creating a net spin on a triangle, modes $\alpha_x$, $\alpha_y$, and $\beta_0$ increase its exchange energy $J(\s_1+\s_2+\s_3)^2/2$. These modes are therefore hard. The remaining modes $\beta_x$, $\beta_y$, and $\alpha_0$ are soft.

The addition of anisotropies harden the soft modes \cite{Chen20} by introducing finite corrections to their energies at $\mathbf{k} = 0$. The $\bm{\beta}$ doublet is lifted from zero energy by a combination of the DM interaction and an easy-plane anisotropy, separating it from the $\alpha_0$ mode. Further, a local easy-axis anisotropy characterized by ($\delta$) gaps the $\alpha_0$ singlet ($\sim \sqrt{\delta^3/J}$) and splits the $\bm{\beta}$ doublet ($\sim \delta/J$) making the two modes non-degenerate at the $\Gamma$ point, see Eq.~(\ref{eq.alpha_gap}), Eq.~(\ref{eq.beta_gap}), and details in Appendix.~\ref{sec.app.gaps}.

However, since in most situations $\lbrace\delta,D\rbrace /J \ll 1 $, for example in Mn$_3$Ge see Table.~\ref{tab: exchange}, we can safely drop this soft mode hardening effect in our theory. This assumption allows us to integrate out the hard modes $\bm\alpha$, and $\beta_0$ to obtain a theory in terms of soft modes only.

\subsection{Field theory for the soft modes}
Here we outline the spin wave field theory for the generic hexagonal antiferromagnet. The kinematic term, like in the case of the two sublattice antiferromagnet Eq.~(\ref{eq.kin-L-1}), originates from the local Berry phase Eq.~(\ref{eq.L-B}). For a spin confined to the $xy$ plane $\theta \simeq \pi/2$, and hence $(\cos\theta - 1)~\dot{\phi} \simeq~(\pi/2-\theta_i) ~ \dot{\phi_i}$ for each sublattice $i = 1,2,3$. For the triangle this leads to a dynamical term, expressed in terms of the normal modes:

\begin{equation}
\mathcal L_{B}
= \mathcal S \sum_{n=1}^3 (\pi/2-\theta_n)  \dot{\phi}_n
= \mathcal S (\dot{\alpha}_0 \beta_0 - \bm{\alpha}\cdot\dot{\bm{\beta}}),
  \label{eq:spin-kinetic}
\end{equation}
where $\mathcal S$ is the spin density on a single sublattice. From this form of the Lagrangian we can see that $\mathcal S\beta_0$ serves as the canonical momentum for $\alpha_0$, whereas $-\mathcal S\bm\alpha$ is the momentum conjugate to $\bm \beta$, as anticipated in Sec. \ref{sec:hexagonal-geometry-canonical}.

It is convenient to rewrite the energy of nearest-neighbor exchange interactions (\ref{eq.Heisenberg-afm}) in terms of the net spin of a magnetic unit cell, 
\begin{equation}
    U = \frac{J}{2}(\mathbf{S}_1 + \mathbf{S}_2 + \mathbf{S}_3)^2 - \frac{3 J S^2}{2}.
\end{equation}
From this we obtain the energy density to the zeroth order in the spatial gradients, which includes only the hard modes,

\begin{equation}
\mathcal U =  \frac{\mathcal A}{2} (\bm \alpha \cdot \bm \alpha + 2 \beta_0^2).
\label{eq:U-no-gradients}
\end{equation}
Here $\mathcal A$ is a lattice-dependent constant proportional to $J S^2$. The Lagrangian density now reads:
\begin{equation}
\mathcal{L} = \mathcal S (\dot{\alpha}_0 \beta_0 - \bm{\alpha}\cdot\dot{\bm{\beta}}) - \frac{\mathcal A}{2}(\bm \alpha \cdot \bm \alpha + 2 \beta_0^2) 
\end{equation}
The equations of motion for the hard modes,
\begin{equation}
\mathcal S \dot{\bm \beta} = -\mathcal A \bm \alpha,
\quad
\mathcal S \dot{\alpha}_0 = 2\mathcal A \beta_0,
\end{equation}
can be used to integrate them out and in the process to generate a kinetic energy for the soft modes:
\begin{equation}\label{eq.inertia_triangle}
\mathcal L_{\text{kin}} = \frac{\rho_\alpha}{2} \dot{\alpha}_0^2 + \frac{\rho_\beta}{2} \dot{\bm\beta} \cdot \dot{\bm\beta}. 
\end{equation}
Here $\rho_\alpha = \mathcal S^2/2\mathcal A$ and $\rho_\beta = \mathcal S^2/\mathcal A$ are inertia densities for the soft modes $\alpha_0$ and $\bm\beta$.

Exchange energy of the soft modes vanishes at the zeroth order in the gradient expansion because uniform $\alpha_0$ and $\bm\beta$ represent global spin rotations. The lowest nonvanishing contributions to the exchange energy come at the second order in the gradient expansion. The form of these second-order terms is strongly constrained by the hexagonal or trigonal symmetry of the lattice. We discuss it next for the singlet $\alpha_0$ and the doublet $\bm \beta$.

\subsubsection{Singlet}
The singlet mode $\alpha_0$ has a simple theory. Its Lagrangian density consists of a kinetic energy with mass density $\rho_\alpha$ and a potential energy quadratic in the gradients of $\alpha_0$:  
\begin{equation}\label{eq:L-alpha0}
\mathcal L = 
    \frac{\rho_\alpha}{2}  \dot{\alpha}_0^2 
    - \frac{\kappa}{2}  
        \partial_i \alpha_0 \,
        \partial_i \alpha_0.
\end{equation}
The stiffness $\kappa$ is determined by exchange interactions. Summation is assumed over doubly repeated Cartesian indices $i = x,y$. 
As often happens in highly symmetric solids, the effective Lagrangian (\ref{eq:L-alpha0}) obeys not just the discrete symmetries of the point group $D_3$ but also the full rotational symmetry SO(2). Spin waves have a linear dispersion $\omega = v k$ with the speed $v = \sqrt{\kappa/\rho_\alpha}$.

\subsubsection{Doublet}
The continuum theory for the doublet is more involved as the doublet field $\bm \beta$ itself transforms like a vector under spatial rotations in xy space. The Lagrangian of this field has the following form: 
\begin{equation}\label{eq:L-beta}
\mathcal L = 
    \frac{\rho_\beta}{2}  \dot{\beta}_i^2
    - \frac{C_{ijkl}}{2}  \beta_{ij} \beta_{kl}
    - \frac{\tilde{C}_{ijkl}}{2}  
        \tilde{\beta}_{ij} \tilde{\beta}_{kl}.
\end{equation}
Here we have introduced symmetrized and anti-symmetrized gradients, 
\begin{equation}
\beta_{ij} \equiv 
    \frac{1}{2}
        (\partial_i \beta_j 
        + \partial_j \beta_i),
\quad
\tilde{\beta}_{ij} \equiv 
    \frac{1}{2}
        (\partial_i \beta_j 
        - \partial_j \beta_i).
\end{equation}
The inertia density $\rho_\beta$ is generally different from its counterpart $\rho_\alpha$ for the singlet mode. The stiffness coefficients, determined by the exchange interactions, are fourth-rank tensors with the following symmetry properties: $C_{ijkl}$ is symmetric and $\tilde{C}_{ijkl}$ is antisymmetric under the exchanges $i \leftrightarrow j$ and $k \leftrightarrow l$; both tensors are symmetric under the exchange $(ij) \leftrightarrow (kl)$.

The structure of the Lagrangian (\ref{eq:L-beta}) is highly reminiscent of the theory of elasticity in two dimensions. Here $\beta_i$ identifies with the lattice displacement, $\beta_{ij}$ with strain, and $\tilde{\beta}_{ij}$ with rotation of the lattice. In a solid, rotations do not increase the elastic energy, so $\tilde{C}_{ijkl} = 0$ for lattice vibrations. For spin waves, $\tilde{C}_{ijkl} \neq 0$ in general.

As with the elastic constants, the highly symmetric hexagonal environment drastically reduces the number of independent potential coefficients. Both fourth-rank tensors can be expressed in SO(2)-invariant forms: 
\begin{eqnarray}\label{eq:C-isotropic-solid}
C_{ijkl} &=& 
    \lambda \delta_{ij} \delta_{kl}
    + \mu (\delta_{ik} \delta_{jl}
            + \delta_{il} \delta_{jk}),
\nonumber\\
\tilde{C}_{ijkl} 
    &=& \tilde{\mu} \epsilon_{ij} \epsilon_{kl}
    = \tilde{\mu} (\delta_{ik} \delta_{jl}
            - \delta_{il} \delta_{jk}).
\end{eqnarray}
Here $\delta_{ij}$ is the Kronecker delta and $\epsilon_{ij}$ is the antisymmetric Levi-Civita symbol, $\epsilon_{xy} = - \epsilon_{yx} = +1$. The Lam{\'e} parameters $\lambda$ and $\mu$ determine the bulk modulus $\lambda + \mu$ (in 2 dimensions) and the shear modulus $\mu$. To continue the analogy with a solid, we will refer to $\tilde{\mu}$ as the rotation modulus. The explicit form of the Lagrangian for the $\bm\beta$ modes is
\begin{equation}\label{eq:L-beta-explicit}
\mathcal L = 
    \frac{\rho_\beta}{2}  \dot{\beta}_i^{2} 
    - \frac{\lambda}{2} \, 
    \partial_i \beta_i \, \partial_j \beta_j
    - \frac{\mu + \tilde{\mu}}{2} 
        \partial_i \beta_j \partial_i \beta_j
    - \frac{\mu - \tilde{\mu}}{2} 
        \partial_i \beta_j \partial_j \beta_i.
\end{equation}

Spin waves for the $\bm \beta$ modes with longitudinal and transverse polarizations have the propagation speeds 
\begin{equation}\label{eq:speed-of-spin-waves}
v_{||} = \sqrt{\frac{\lambda+2\mu}{\rho_\beta}}, 
\quad
v_{\perp} = \sqrt{\frac{\mu+\bar{\mu}}{\rho_\beta}}.
\end{equation}

\subsubsection{Six-fold symmetric gradient}
The continuum spin-wave Lagrangians (\ref{eq:L-alpha0}) and (\ref{eq:L-beta-explicit}) exhibit full $SO(2)$ rotational invariance. In a hexagonal solid, this symmetry is only approximate and is explicitly broken if we include terms of higher orders in the gradients. These higher order terms are constrained by the $D_3$ point-group symmetry.

The $D_3$ symmetry allowed terms can be constructed from the soft modes as follows. Take three unit vectors $\mathbf n_1$, $\mathbf n_2$, and $\mathbf n_3$ making angles of $120^\circ$ with one another. For arbitrary vectors $\mathbf a$, $\mathbf b$, and $\mathbf c$, the sum
\begin{equation}\label{eq:six-fold-anisotropy}
\sum_{i = 1}^{3}
    (\mathbf a \cdot \mathbf n_i)
    (\mathbf b \cdot \mathbf n_i)
    (\mathbf c \cdot \mathbf n_i) 
\end{equation}
is invariant under $120^\circ$ rotations. Furthermore, the square of this quantity is invariant under $60^\circ$ rotations.

For the scalar $\alpha_0$ mode, the only vector available is the gradient operator $\nabla$ (or the wave-vector $\mathbf k$), so we take $\mathbf a = \mathbf b = \mathbf c = \nabla$. A quantity invariant under $60^\circ$ rotations is 
\begin{equation}\label{eq.L6}
\mathcal L_6 = 
- \frac{\sigma_\alpha}{8}
    \left[
        \left(
            \partial_x^3 - 3 \partial_x \partial_y^2
        \right)
        \alpha_0
    \right]^2.
\end{equation}
Adding this term to the Lagrangian of the $\alpha_0$ mode alters the magnon dispersion, warping the cone $\omega = vk$ as follows: 
\begin{equation}
\omega^2 = v^2 k^2 + \frac{\sigma_\alpha}{\rho_\alpha} k^6 \cos^2{3\phi},    
\end{equation}
where $\phi$ is the angle at which the magnon propagates in the $xy$ plane, $\mathbf k = (k \cos{\phi}, k\sin{\phi})$. The warping is strongly suppressed near the center of the Brillouin zone. 

For the $\bm \beta$ mode which transforms as a vector under rotations in the xy plane, we have two in plane vectors available for our construction, $\nabla$ and $\bm \beta$. The relevant invariant is 
\begin{equation}
\mathcal L_6 = 
- \frac{\sigma_\beta}{2}
    \left[
        (\partial_x^2 - \partial_y^2) \beta_x
        - 2 \partial_x \partial_y \beta_y
    \right]^2.
\end{equation}
For nondegenerate longitudinal and transverse modes ($v_{||} \neq v_\perp$), the magnon dispersions are warped as follows: 
\begin{eqnarray}
\omega^2 &=& v_{||}^2 k^2 
    + \frac{\sigma_\beta}{\rho_\beta} 
        k^4 \cos^2{3\phi}, 
\nonumber\\
\omega^2 &=& v_{\perp}^2 k^2 
    + \frac{\sigma_\beta}{\rho_\beta} 
        k^4 \sin^2{3\phi}.
\end{eqnarray}
The warping for the $\bm \beta$ modes comes at a lower order in the gradient expansion and is therefore more pronounced than for the $\alpha_0$ mode. Note that if either of the velocities $(v_{||},v_{\perp})$ are zero this makes the six-fold pattern very prominent for that mode. 

\section{Familiar Examples}\label{sec.ped_examples}
Let us now explicitly construct the field theory for the cases of the nearest neighbor triangular antiferromagnet and the kagome antiferromagnet, see Fig.~\ref{fig:lattice}. The difference between the two is the coordination number of each site. For the triangular lattice each site has a coordination number of six while for the kagome the coordination number is four. This affects the spin density and the gradient expansions which have to be calculated separately for each type of lattice.

\begin{figure*}[t]
     \includegraphics[width=2\columnwidth]{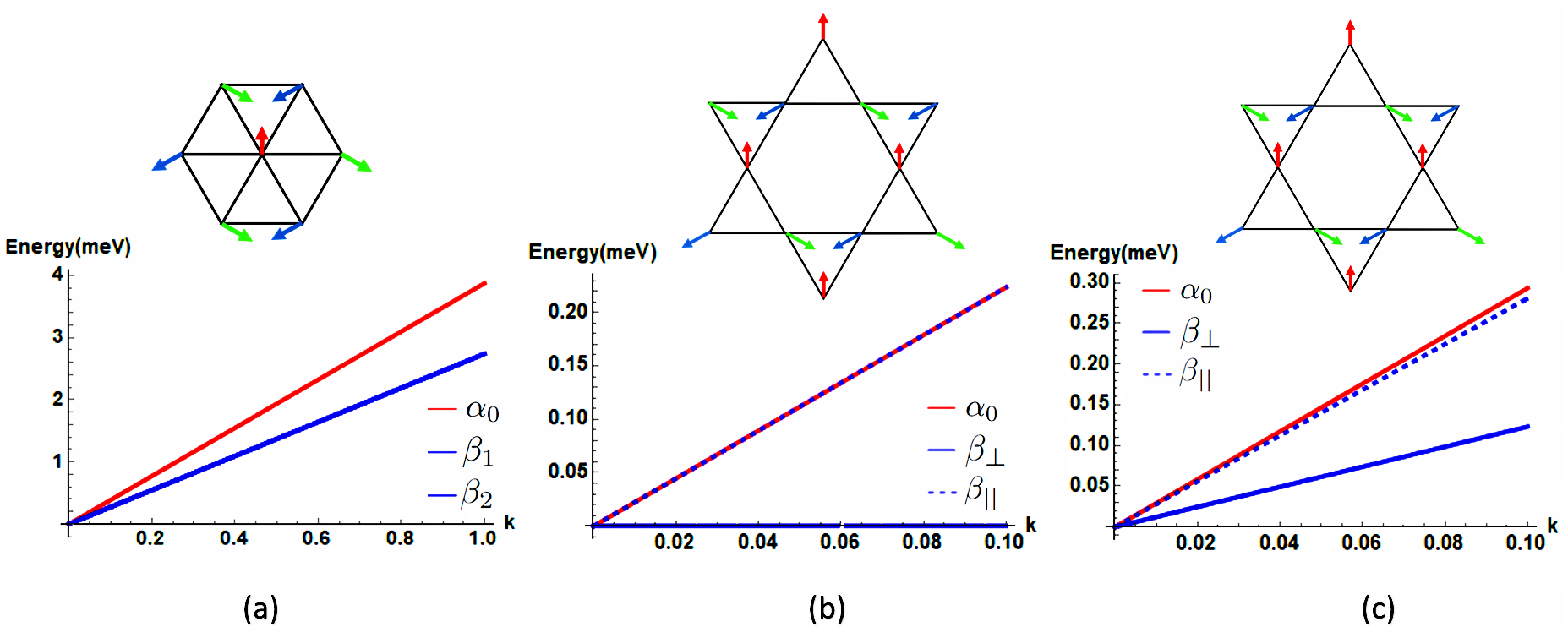}
        \caption{Dispersion $\omega(k)$ of the three Goldstone modes for hexagonal antiferromagnets with nearest-neighbor exchange $J = 10$ meV. (a) Triangular lattice with nearest-neighbor exchange only. (b) Kagome lattice with nearest-neighbor exchange only. (c) Kagome lattice with ferromagnetic next-nearest-neighbor exchange  $J_\text{nnn} = -0.5~$meV.}
        \label{fig:example-dispersions}
\end{figure*}

For any individual lattice system we start with the kinetic energy derived in Eq.~(\ref{eq.inertia_triangle}). The inertia for the soft modes $\rho_\alpha$ and $\rho_\beta$ need to be determined for each lattice. For the soft mode contribution to the potential energy density $\mathcal{U}$ we do a gradient expansion of the exchange interaction in the soft fields with the amplitudes of the hard modes set to zero. This is combined with the kinetic energy density to form the full Lagrangian density $\mathcal{L} = \mathcal{L}_{\text{kin}} - \mathcal{U}_g(\nabla\alpha_0,\nabla\beta_x,\nabla\beta_y)$.

\subsection{Triangular antiferromagnet}
In the nearest neighbor Heisenberg model on the triangular lattice \cite{Dombre1989, Chubukov1999}, the magnetic unit cell has the area $A = (3\sqrt{3}/2)a^2$ where $a$ is the nearest neighbor distance, see Fig \ref{fig:lattice}(b). The spin density is then : $\mathcal{S} = 2S/(3\sqrt{3}a^2)$ and the energy density parameter $\mathcal{A} = (\sqrt{3}JS^2)/a^2$ This results in the inertia:
\begin{equation}
\rho_{\beta} = \frac{\mathcal{S}^2}{\mathcal{A}} = \frac{4}{27\sqrt{3}~Ja^2} = 2\rho_{\alpha}.
\end{equation}
In addition gradient expansion in the soft modes yields the energy density:
\begin{equation}
    \mathcal{U}_g = \frac{J S^{2}}{4\sqrt{3}} \left[(\nabla\alpha_{0})^{2} + (\nabla\beta_{x})^{2} + (\nabla\beta_{y})^{2}\right],
\end{equation}
One can identify the constants $\kappa = J S^{2}/(2\sqrt{3})$ for the $\alpha_0$ singlet and for the $\bm{\beta}$ doublet $\lambda = 0$, and $\mu = \tilde{\mu}= J S^{2}/(4\sqrt{3})$.

The $\alpha_0$ mode has the speed $v = \frac{3\sqrt{3}}{2} JSa$. The $\bm \beta$ modes are degenerate and have speeds $v_{||} = v_\perp = v/\sqrt{2}$, see Eq.~(\ref{eq:speed-of-spin-waves}) and Fig.~\ref{fig:example-dispersions}(c). The degeneracy is associated with the special values of the Lam{\'e} coefficients, $\lambda = 0$ and $\mu = \tilde{\mu}$, and reflects a higher, $\text{SO}(2) \times \text{SO}(2)$ symmetry of the Lagrangian, 
\begin{equation}
\mathcal L = 
    \frac{1}{2} \rho_\beta \dot{\beta}_i \dot{\beta}_i   
    - \mu \, \partial_i \beta_j 
        \, \partial_i \beta_j,
\end{equation}
where one $\text{SO}(2)$ rotates spatial coordinates and the other transforms components of the $\bm \beta$ doublet.

\subsection{Kagome antiferromagnet}
For the nearest-neighbor kagome antiferromagnet \cite{Harris1992} the magnetic unit cell area in $ A = (2\sqrt{3})a^2$. The spin density is given by $\mathcal{S} = S/(2\sqrt{3}a^2)$, see Fig.~\ref{fig:lattice}(b). This gives the energy density parameter $\mathcal{A} = (\sqrt{3} J S^2)/(2 a^2)$. From this we can extract the inertia for the two modes:
\begin{equation}
    \rho_{\beta} = \frac{\mathcal{S}^2}{\mathcal{A}} = \frac{1}{6\sqrt{3}Ja^2} = 2\rho_{\alpha}.
\end{equation}
The soft mode expansion of the exchange interaction yields the following energy density:
\begin{equation}\label{eq.U_exch_Kag_J}
\mathcal{U}_g = \frac{J S^{2}}{8\sqrt{3}}\left[ (\nabla \alpha_0)^{2} + 2 (\nabla \cdot \bm \beta)^{2}  \right].
\end{equation}
The constants for the kagome lattice are hence $\kappa = J S^{2}/4\sqrt{3}$ for the $\alpha_0$ singlet and for the $\bm{\beta}$ doublet $\lambda = J S^{2}/2\sqrt{3}$, and $\mu = \tilde{\mu}= 0$.

The $\alpha_0$ mode and the longitudinal part of the $\bm \beta$ mode have the speed $v_{\alpha} = v_{||} =\sqrt{3}JSa$, whereas the transverse $\bm \beta$ mode has $v_\perp = 0$, where $a$ is the nearest neighbor distance, see Fig.~\ref{fig:example-dispersions}(a)\cite{Harris1992}. The zero transverse speed is associated with the vanishing shear and rotation moduli, $\mu = \tilde{\mu} = 0$ in the dual elasticity theory. In this sense, the nearest-neighbor kagome antiferromagnet resembles a fluid. Adding exchange interactions beyond nearest neighbors generates a finite shear stiffness and a nonzero speed for the transverse $\bm\beta$ mode. It also lifts the degeneracy of the $\alpha_0$ and longitudinal $\bm\beta$ modes. See Fig. \ref{fig:example-dispersions}(c).

The zero mode that persists throughout the Brillouin zone in the nearest-neighbor kagome antiferromagnet with 120$^\circ$ order does not rise out of a spontaneously broken symmetry. This zero mode is due to an accidental degeneracy. It is possible to rotate any two spins in each triangle about the other spin as the axis of rotation, along a row, at no energy cost. These modes have been seen in experiment \cite{matan06}. If we look at the zero mode shown there it is exactly the $\beta_{\perp}$ mode for a spin wave propagating in the $y$ direction.

This fluid-like behavior in an antiferromagnet has a direct analogy to the continuum elasticity theory of the kagome lattice with nearest-neighbor interactions, which is critical according to the Maxwell criteria for stability \cite{maxwell_elastic1864,Sun12369}. This mechanical system is unstable to distortions with zero modes comprised of twisted triangles along certain directions. These zero gain a sound velocity by an addition of elastic coupling between further neighbors \cite{Mao_PRE_2011_kag_elas}. 

Similarly, for the spin system the addition of further neighbor exchanges, lifts the degeneracy between the $\alpha_0$ and the longitudinal $\beta$ mode and generates a finite velocity for the transverse $\beta$ mode \cite{Harris1992}, see Fig.~\ref{fig:example-dispersions}(b).

\section{Stacked Kagome}\label{sec.Stacked_kagome}
We shall now look into the spin waves for Mn$_3$Ge. This is a layered kagome system where the two layers are displaced relative to each other such that the up triangles of one layer coincide with the down triangles of the layers above and below it. The kagome spin lattice in each plane is comprised of three spin sublattices and have the 120$^\circ$ antivortex magnetic order. Like the planar kagome system the magnetic order here is defined within a single triangle and does not vary along the c-axis. The dominant energy scale is the nearest neighbor, in the kagome plane, antiferromagnetic exchange of strength $J_2$.

Inelastic neutron scattering data, shown in Fig. 4 of \cite{Chen20}, reveal spin waves with high propagation speeds and no evidence of zero modes, which indicates the presence of further-neighbor exchange interactions. The nearest additional interaction that produces this dispersion for the stacked kagome is an interlayer interaction.

In addition, the system has a DM interaction with a $\mathbf{D} = D\hat{z}$ vector that points out of the ab plane. This locks the spins into an antivortex order and minimizes spin canting out of the kagome planes. There is a small on-site easy-axis anisotropy which cants the spins in plane, out of the 120$^\circ$ order, characterized by $\delta$ \cite{Tomiyoshi1982,Brown_1990,Nakatsuji:2015} see Appendix.~\ref{sec.app.ferro.mom}. This energy scale is three orders of magnitude smaller than any exchange energy scale $\delta \ll (J_2,J_4)$. This is evident from our fits to spin wave dispersion data in \cite{Chen20}, see Table.~\ref{tab: exchange}

Thus, we have on our hands a stacked kagome antiferromagnet, where the ordered state hosts the same Goldstone modes as the single layered trigonal lattice antiferromagnets. We apply our theory to this system, extracting analytical expressions for the long wavelength spin-wave velocities and the gaps in the Goldstone modes at $\mathbf{k} = 0$. We use these expressions to fit inelastic neutron scattering data and extract the parameters for the spin Hamiltonian \cite{Chen20}:

\begin{eqnarray}\label{eq:H-mnx}
\mathcal{H}_{JD\delta}=&& \sum_{<i,j>} J_{ij}~ \mathbf{S}_i\cdot \mathbf{S}_j + \sum_{<i,j>} \mathbf{D}_{ij} \cdot (\mathbf{S}_i \times \mathbf{S}_j)
\nonumber\\
&-& \sum_{i} \delta(\hat{\mathbf{n}}_i \cdot \mathbf{S}_i)^2.
\end{eqnarray}
Here $JD\delta$ stands for a model containing Heisenberg exchanges, collectively $J$, a DM interaction $D$ and local anisotropy $\delta$. The local anisotropy rises from an easy-axis at each Mn site $i = 1,2,3$. The axis is directed towards the nearest Ge site, represented by the unit vectors $\mathbf{\hat{n}}_i$ \cite{Nakatsuji:2015,Chen20}.

The bi-layer unit forms a David's star motif consisting of an up triangle in the lower (blue) layer and a down triangle of the upper (red) layer, the central plaquette in Fig.~\ref{fig:interplane-interaction}(a). An effective description of the system requires two sets of modes: $(\alpha_{0}^A,\bm{\alpha}^A,\beta_{0}^A,\bm{\beta}^A)$ for the A layer and $(\alpha_{0}^B,\bm{\alpha}^B,\beta_{0}^B,\bm{\beta}^B)$ for the B layer. The theory is better expressed in terms of symmetric and antisymmetric combinations of the two sets:
\begin{equation}
\zeta = \frac{\zeta^A + \zeta^B}{\sqrt{2}}, ~~~ \bar{\zeta} = \frac{\zeta^A - \zeta^B}{\sqrt{2}},
\end{equation}
where $\zeta$ stands for any of the $\alpha$ or $\beta$ modes. We shall derive the field theory for the stacked kagome system in terms of these twelve modes.

\subsection{Kinetic term and inertia}
The net Berry phase, Eq.~(\ref{eq:spin-kinetic}) for each layer, can be expressed in terms of the symmetric and antisymmetric modes:

\begin{equation}\label{eq.L_B}
\mathcal{L}_{B} = 
\mathcal S 
(\dot{\alpha}_0\beta_0 
    -\bm {\alpha}\cdot \dot{\bm {\beta}}  + \dot{\bar{\alpha}}_0\bar{\beta_0} - \bm {\bar{\alpha}}\cdot \dot{\bm {\bar{\beta}}}).
\end{equation}

\begin{figure}[t]
    \includegraphics[width=\columnwidth]{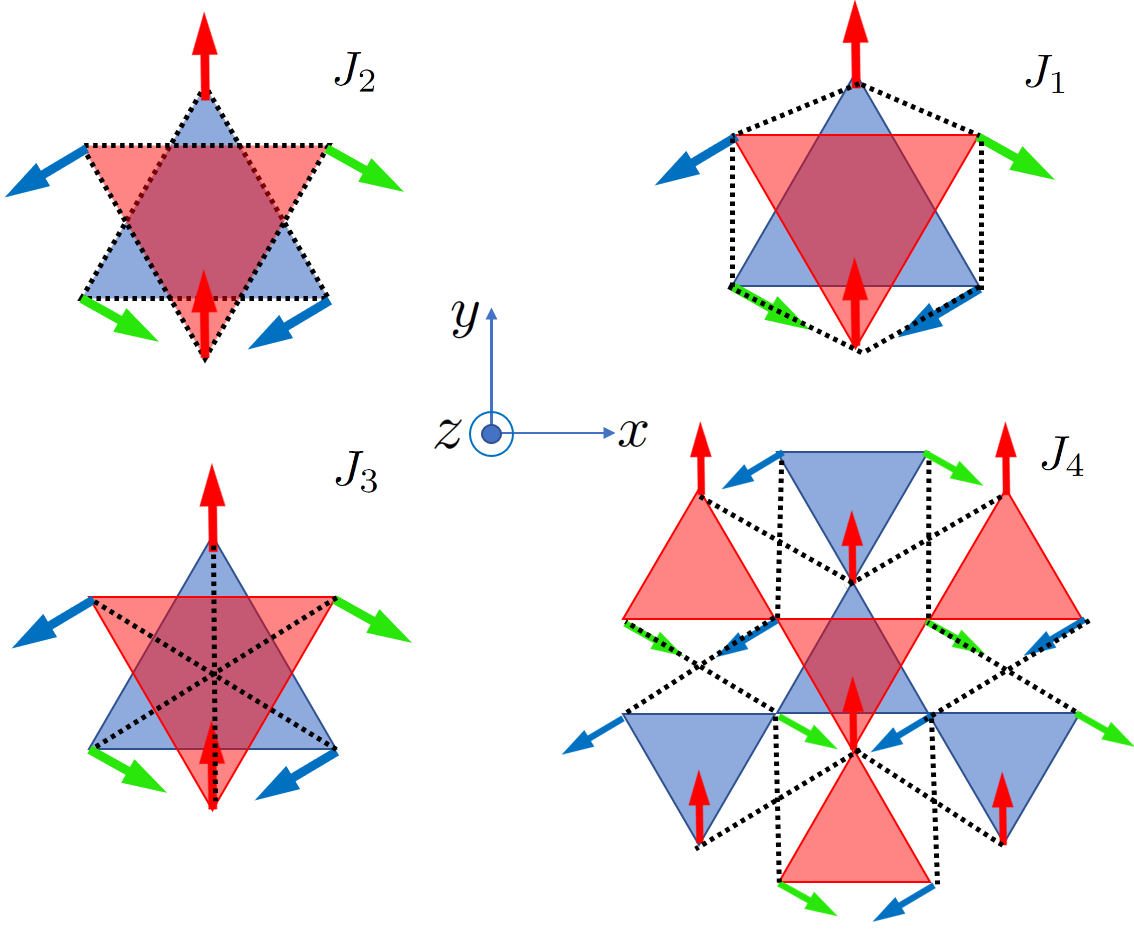}
    \centering
    \caption{Heisenberg exchange interactions in Mn$_3$Ge, shown as dashed lines: intralayer exchange $J_2$ and interlayer exchanges $J_1$, $J_3$, and $J_4$.}
    \label{fig:interplane-interaction}
\end{figure}

Here $\mathcal{S} = S/V$ where $V = (4\sqrt{3})a^2c$ is the volume of the magnetic unit cell, $c$ is the AB layer separation. For the potential energy we have to consider three types of exchange interactions, see Fig.~\ref{fig:interplane-interaction}. The dominant exchange is the intralayer nearest neighbor antiferromagnetic exchange of strength $J_2$. To reproduce the isotropic dispersion seen in \cite{Chen20} we add interlayer couplings $J_{1}$ and $J_{4}$.  The index $i$ in $J_i$ labels the $i$th nearest neighbor.

Fits to the spin-wave data reveal the values of $J_1$ and $J_3$ to be much smaller than $J_2$ and $J_4$. In fact, to the first order the spin-wave dispersion depends on the sum $J_1+J_3$. We retain only one of these couplings, $J_1$, and set $J_3=0$. A nonzero $J_1$ gives rise to some interesting features such as an anisotropic dispersion dispersion of spin waves at small $\mathbf k$. $J_4$ is the nearest exchange that produces an isotropic dispersion for the flat $\beta_\perp$ band. 

As before, we can convert the Berry phase into a kinetic energy by integrating out the hard modes. In this case there are six such modes. For the examples we worked out in Sec. \ref{sec.ped_examples}, the fields we retained were the ones that were soft under exchange. We perform the same exercise here but with a bit more scrutiny. The energy density $\mathcal{U}$ obtained from expansion of three exchange interactions:

\begin{eqnarray}\label{eq.U_ex_hard}
\mathcal U &=& C_1\left(\bm{\alpha}^{2} + 2 \beta_{0}^{2}\right) + C_2 \bm{\bar{\alpha}}^{2} + C_3 \bar{\beta}_{0}^{2}\\ \nonumber
&+& C_{4}\left(\bm{\bar{\beta}}^{2} + \bar{\alpha}_{0}^{2}\right) + \mathcal{U}_g,
\end{eqnarray}

where $\mathcal{U}_g$ contains the gradients of the modes. The constants $C_{n}$ are:
\begin{eqnarray}
\label{eq.gaps_gamma}
    C_{1} &=& \left(\frac{\sqrt{3}}{8}J_{2} + \frac{\sqrt{3}}{8}J_1 \right)\frac{S^2}{c a^2}. \\ \nonumber
    C_{2} &=& \left(\frac{\sqrt{3}}{8}J_2 + \frac{J_1}{8\sqrt{3}} - \frac{J_4}{\sqrt{3}} \right)\frac{S^2}{c a^2}. \\ \nonumber
    C_{3} &=& \left(\frac{\sqrt{3}}{4}J_2 - \frac{J_1}{4\sqrt{3}} -  \frac{J_4}{\sqrt{3}}\right)\frac{S^2}{c a^2}. \\ \nonumber \vspace{6mm}
    C_{4} &=& \left(\frac{J_1}{2\sqrt{3}} - \frac{J_4}{\sqrt{3}} \right)\frac{S^2}{c a^2}. \vspace{6mm}
\end{eqnarray}
In the presence of the interlayer exchanges $J_1$ and $J_4$, all the antisymmetric modes pick up zeroth order in gradient energy contributions. Three gapless modes (Goldstones) remain: the symmetric modes $(\alpha_0,\bm {\beta})$.

The interlayer couplings can cause instabilities (negative gap energies) in the 120$^{\circ}$ order if we have a ferromagnetic (antiferromagnetic) exchange between sites of the opposite (same) sublattice. Here, for instance, if $\text{sgn}(J_1)<0$ or $\text{sgn}(J_4)>0$, then we have the unstable situation $C_4 < 0$. For the experiment \cite{Chen20} the fits require an antiferromagnetic $J_1$ and a ferromagnetic $J_4$. This provides positive energies at the zeroth order in gradients to all the antisymmetric modes and there are no instabilities. The full theory with all twelve modes is:
\begin{equation}
\mathcal{L} = \mathcal{L}_B - \mathcal{U},
\end{equation}
where $\mathcal{U}$ is defined in Eq.~(\ref{eq.U_ex_hard}). From this we can integrate out six modes $(\beta_0,\bar{\beta}_0,\bm{\alpha},\bm{\bar\alpha})$ using their equations of motion. These modes are hard due to $J_2$ and hence their gradients are not considered in $\mathcal{U}_g$. This results in a theory:
\begin{eqnarray}\nonumber
\label{eq.gamma_point_mnx}
    \mathcal{L} &=& \frac{\rho_{\alpha}}{2}\dot{\alpha}_0^2 + \frac{\rho_{\beta}}{2}\dot{\bm{\beta}}^2 +  \frac{\rho_{\bar\alpha}}{2}\dot{\bar\alpha}_0^2 + \frac{\rho_{\bar\beta}}{2}\dot{\bar{\bm{\beta}}}^2 \\
    &-& 
    C_{4}\left(\bm{\bar\beta}^{2} + \bar\alpha_{0}^{2}\right) - \mathcal{U}_g.
\end{eqnarray}
The inertia for the $\alpha_{0}$ and $\bm{\beta}$ modes is generated by integrating out the hard $\beta_{0}$ and $\bm{\alpha}$ modes, respectively:
\begin{equation}\label{eq.inertia-sym}
    \rho_{\beta} = \frac{\mathcal{S}^{2}}{2 C_{1}} = \frac{1}{12\sqrt{3}(J_1 + J_2)a^2 c} = 2\rho_{\alpha}.
\end{equation}
Similarly the inertias for the antisymmetric modes are:
\begin{equation}
    \rho_{\bar\beta} = \frac{\mathcal{S}^2}{2C_2}, ~~ \rho_{\bar\alpha} = \frac{\mathcal{S}^2}{2 C_3}.
\end{equation}
These modes are not critical to our study as they are hard in Mn$_3$Ge, see $C_4$ in Eq.~(\ref{eq.gamma_point_mnx}). This allows us the freedom to drop the space-time gradients of all the antisymmetric fields. The resulting kinetic energy we work with is:
\begin{equation}
\label{eq.kinetic_mnx}
    \mathcal{L}_{\text{kin}} \simeq \frac{\rho_{\alpha}}{2}\dot{\alpha}_0^2 + \frac{\rho_{\beta}}{2}\dot{\bm{\beta}}^2.
\end{equation}

We now calculate the interaction energy density generated by the gradient expansion of the Heisenberg exchanges in the remaining modes $(\alpha_{0},\bm{\beta})$ and $(\bar\alpha_{0},\bm{\bar\beta})$. Though the antisymmetric modes are hard in this problem (due to $J_4$) we do not set their amplitudes to zero in the gradient expansion. This is done to retain terms linear in their gradients.  We proceed one exchange interaction at a time, highlighting the features in each case.

\subsection{Intralayer interactions}

Heisenberg antiferromagnetic exchange between nearest neighbor sites confined to a single kagome plane, $J_2$  (see Fig.~\ref{fig:interplane-interaction}(a)) reproduces the kagome lattice example worked out earlier. This is the dominant exchange term in this compound. The energy density is
\begin{eqnarray}\label{eq.U_exch_J2}
\mathcal{U}_g &=& \frac{S^{2}}{16\sqrt{3}~ c}J_2\left( \nabla\alpha_0^{2} + \nabla \bar{\alpha}_0^{2}  \right) \\ \nonumber
&+& \frac{S^{2}}{8\sqrt{3}~ c}J_2\left[ (\nabla \cdot \bm{\beta})^{2} + (\nabla \cdot \bm{\bar\beta})^{2} \right],
\end{eqnarray}
where $\nabla \equiv (\partial_x, \partial_y)$ includes in-plane gradients only.
In the absence of interlayer coupling, the symmetric and antisymmetric fields are degenerate. This implies that the inertia for the symmetric and antisymmetric modes is the same, $\rho_{\alpha} = \rho_{\bar\alpha}$ and $\rho_{\beta} = \rho_{\bar\beta}$.

For the $\alpha_0$ modes we have, $\kappa = J_2 S^{2}/(8\sqrt{3}c)$, and hence $v_{\alpha_0} = \sqrt{3} J_2 Sa$. For the ${\bm \beta}$ modes from elasticity theory we can read off the elasticity moduli:  $\lambda = J_2 S^{2}/(4\sqrt{3}c)$, $\mu = \tilde{\mu} = 0 $ and hence the velocities:
\begin{equation}\label{eq.beta-vel-j2}
v_{\beta_{||}} = \sqrt{3} J_2 S a, ~~  v_{\beta_\perp} = 0.
\end{equation}
The `solid' has zero shear modulus and hence has a flat mode in the direction perpendicular to a propagating elastic wave. Since $\rho_{\beta} = 2\rho_{\alpha}$ the two dispersive modes propagate at the same speed $v_{\alpha} = v_{\beta_{||}}$

\subsection{Interlayer interactions}
To reproduce the dispersion observed in the experiment \cite{Chen20} we need to find exchange interactions that endow the flat $\beta_{\perp}$ mode with an isotropic dispersion. The nearest interaction that does the job is  $J_4$, shown in Fig.~\ref{fig:interplane-interaction}. As indicated before we retain the small $J_1$ coupling and show that although it is ineffective in producing an isotropic quadratic dispersion for $\beta_{\perp}$ it has some interesting features. The gradient contribution to the potential energy density $\mathcal{U}$ from the interlayer exchanges is
\begin{widetext}
\begin{eqnarray}
    \label{eq.U-J}  
    \mathcal{U}_g &=&  \frac{J_1-2J_4}{2\sqrt{3}}\frac{S^2}{a^2c} \left(\bm{\bar\beta}^{2} + \bar\alpha_{0}^{2}\right) - \frac{J_1 + J_4}{6}\frac{S^2}{ac}\left[\bar{\beta}_x (\partial_y \beta_x + \partial_x \beta_y) + \bar{\beta}_y (\partial_x \beta_x - \partial_y\beta_y)\right] \\ \nonumber
     &+& \frac{J_1-8J_4}{48\sqrt{3}}\frac{S^2}{c}\left[(\partial_x\alpha_{0})^{2} + (\partial_y\alpha_{0})^{2}\right]  - \frac{J_4}{8\sqrt{3}}\frac{S^2}{c}(\partial_x \beta_{x} + \partial_y \beta_{y})^{2} + \frac{J_1 - 5 J_4}{24\sqrt{3}}\frac{S^2}{c}(\partial_y \beta_{x} - \partial_x \beta_{y})^{2}
\end{eqnarray}
\end{widetext}
where we have dropped the gradients of the hard antisymmetric modes $\bar{\alpha}_{0}$ and $\bm{\bar\beta}$.

\subsubsection{Lifshitz invariants}
Interlayer interactions generate terms that go beyond the simple elasticity theory. These include antisymmetric Lifshitz invariants $\bar{\beta}_i\partial_j\beta_k - \beta_k\partial_j\bar{\beta}_i$. When the hard field $\bm{\bar\beta}$ is integrated out, these terms give rise to a sixfold anisotropy of the spin-wave dispersion near $\mathbf k = 0$. See Appendix \ref{sec.app.mod.elastic} for details. 

\subsubsection{xy-plane velocities}
In the perturbative regime, where we can integrate out the antisymmetric modes $(\bar{\alpha}_0,\bar{\bm\beta})$ from Eq.~(\ref{eq.U-J}), we list the velocities of all the gapless modes in the presence of both in-plane and out-of-plane interactions:
\begin{eqnarray}\label{eq.velocities_total}
v_{\alpha_0}  &=& aS \sqrt{( J_1 + J_2 )(3 J_2 - 8 J_4 + J_1)},\\ \nonumber
v_{\beta_{||}} &=& aS \sqrt{(J_1 + J_2)(6J_2 - 5J_4 - 2J_1 - 9J_{14})/2}, \\ \nonumber
v_{\beta_{\perp}} &=& 3aS \sqrt{(J_1 + J_2)\left( -J_4  - J_{14}\right)/2},
\end{eqnarray}
where $J_{14} = J_1 J_4/(J_1-2 J_4)$.

Note that the transverse mode $\beta_\perp$ acquires a nonzero speed $v_{\beta_{\perp}}$ only if $J_4 \neq 0$. The interaction $J_1$ alone (or, equivalently, $J_3$) also lifts the $\beta_\perp$ mode from zero frequency but does so in a rather anisotropic manner, with the frequency staying zero along certain directions. See  Fig.~\ref{fig:dispersions}(a). 

\subsubsection{Out-of-plane velocities}
The out of plane dispersions for the $\alpha_0$ mode and $\bm{\beta}$ mode are given by:
\begin{equation}
    \rho_{\alpha}\omega_{\alpha_0}^{2} = \rho_{\beta}\omega_{\beta}^{2} = \left(\frac{J_1}{4\sqrt{3}} - \frac{J_4}{2\sqrt{3}}\right) \frac{c}{a^2}k_z^2,
\end{equation}
with $\rho_{\alpha}$ as given by Eq.~(\ref{eq.inertia-sym}) and c is the interlayer separation.The out-of-plane spin-wave velocities are
\begin{equation}
v_{\alpha_0}^z = \sqrt{2}v_{\beta}^z = cS \sqrt{6(J_1 + J_2)\left(J_1 - 2J_4\right)}.
\end{equation}

\subsubsection{Energy gaps}
The anisotropy terms in Eq.~(\ref{eq:H-mnx}) are a DM interaction, characterized by the DM vector $\mathbf{D} = D\hat{\mathbf{z}}$ and an easy-axis anisotropy, of strength $\delta$, where the local easy axis at an Mn site point towards the nearest Ge site \cite{Nakatsuji:2015,Chen20}. The easy axis breaks the $O(2)$ symmetry in the $xy$ plane and as a result lifts the $\alpha_0$ mode to a finite energy.
\begin{equation}
\label{eq.alpha_gap}
     E_{\alpha} = 3S \sqrt{\frac{2\delta^{3}}{J_1+J_2}}.
\end{equation}
The soft $\bm\beta$ doublet are sensitive to both the DM interaction and local anisotropy. To the lowest order in $D/J$ and $\delta/J$, the doublet acquires an energy gap
\begin{equation}
\label{eq.beta_gap}
    E_{\bm\beta} =  S\sqrt{3(J_1 + J_2)(2\sqrt{3}D + \delta)}.
\end{equation}
At a higher order in the local anisotropy, the doublet is split:
\begin{equation}
\label{eq.beta_split}
    \frac{\Delta E_{\bm\beta}}{E_{\bm\beta}} = \frac{\delta}{6 (J_1 + J_2)} \frac{4\sqrt{3}D - \delta}{2\sqrt{3}D + \delta }.
\end{equation}
The velocity and the gap expressions were used to fit the inelastic neutron data and extract the parameters of the model in Eq. (\ref{eq:H-mnx}), shown in Table \ref{tab: exchange}. The details and particulars of the fitting are discussed in Ref. \onlinecite{Chen20}.

\begin{table}[]
\begin{tabular}{|c|c|c|c|c|c|}
\hline
                                                                   & $J_1S^2$ & $J_2S^2$ & $J_4S^2$ & $DS^2$       & $\delta S^2$                          \\ \hline
\begin{tabular}[c]{@{}c@{}}refined value\\      (meV)\end{tabular} & 0(6)     & 34(7)    & $-17(5)$   & 0.02(1) & $\leq$0.01 \\ \hline
\end{tabular}
\caption {Microscopic parameters of the spin Hamiltonian refined in our work for Mn$_3$Ge. A positive (negative) sign for the exchange parameters corresponds to AFM (FM) interactions. Note that $J_1$ and $J_4$ are inter-plane interactions (see Fig.\ref{fig:interplane-interaction}), while $J_2$, D and $\delta$ are intra-plane interactions.}
\label{tab: exchange}
\end{table}

\section{Conclusion}\label{sec.conclusion}

We have presented a field theory for spin waves in a hexagonal antiferromagnet with three magnetic sublattices and local $D_3$ symmetry in terms of their normal modes. The zero net spin condition imposed on each triangular plaquette leads to a spin wave theory which has three Goldstone modes each with a different velocity, in the generic case. The theory decomposes into a field theory for a singlet $\alpha_0$ and a doublet $\bm{\beta}$. The theory for the doublet maps to a continuum theory for elasticity with the spin wave velocities as `sound' velocities.

We use the familiar settings of the Heisenberg antiferromagnet on the triangular and kagome lattice to demonstrate the features of the field theory. In this case, the two examples are slight outliers because of their highly symmetric lattice environment. 

The triangular lattice has the $\bm{\beta}$ modes as degenerate, and in the kagome we have a degeneracy between the $\alpha_0$ singlet and one of the $\bm{\beta}$ modes while the other one is zero throughout the Brillouin Zone, see Fig.~\ref{fig:example-dispersions}.
We show that the flat mode of the kagome can be anticipated from the elasticity analogy: the mechanical kagome lattice (phonons) with nearest neighbor interaction has zero shear and this property is manifest in our spin wave analog as the flat mode.

Although the spin wave analyses around the 120$^\circ$ ground state of both the triangular Heisenberg antiferromagnet and the kagome antiferromagnet are well documented \cite{Dombre1989,Chubukov1999} their description in terms of three sub lattice field theory is absent from the literature to the best of our knowledge. Additionally, in the case of a local $D_3$ symmetric environment we provide a generic construction scheme for sixfold symmetric terms. This is particularly useful in presence of local anisotropies which break the $O(2)$ symmetry in the plane but keep the sixfold symmetry intact.

We use this theory to describe the spin wave spectrum of Mn$_3$Ge, which has two in-equivalent kagome layers. The analytical expressions for the spin waves and the gaps are used to extract the parameters of the Mn$_3$Ge Hamiltonian.

The study of the normal modes and their natures reveal effective ways of coupling to the magnetic order. External probes like magnetic fields couple to the spins locally, or the net spin of the plaquette and engender terms which are $D_3$ symmetric. These couplings are expressed in the basis of the normal modes, which represent the spin degrees of freedom. Given that the normal modes are $D_3$ symmetric by construction and decouple into a pair of singlets and a pair of doublets we can limit the terms that can be produced based on symmetry properties alone.

For instance, for an external magnetic field the Zeeman coupling is between two time reversal odd vectors: the magnetic field $\mathbf{B}_{\text{ext}}$ and a net spin per plaquette. The only vectors available at the linear order in fields, which are also time reversal odd are, $\mathbf{B}_{\text{ext}}$, and $\bm{\alpha}$. Hence the Zeeman term will be of the form $\mathbf{B}_{\text{ext}}\cdot(\mathcal{R}\bm{\alpha})$ where $\mathcal{R}$ is a 2-d rotation matrix, which accounts for the global $O(2)$ freedom of the spins in the xy plane, see Appendix.~\ref{sec.app.ferro.mom} for details.

Since the magnetism in these materials is intricately linked to the conduction bands of the electrons, through an $s$-$d$ coupling \cite{Liu:2017}, certain features like the location of the Weyl points and, the magnitude of the anomalous Hall response \cite{Liu:2017,Nakatsuji:2015} can be manipulated through the local magnetic order. This is a promising avenue of future work in these materials.

The emergent elasticity theory is also interesting from a more general point of view than just the present scenario, allowing a comparison of this case with other emergent elasticity theories like in skyrmion crystals\cite{Petrova2011skx}.  It also leaves open avenues of investigation along the lines of the duality theory developed in \cite{Pretko2018dual1} and \cite{Pretko2018symenrichdual}, especially since in \ce{Mn3Ge} the non-collinear ground state allows a spin-phonon coupling, which might make a melting transition particularly interesting. 

A detailed study of the soft modes, as provided here, is of use in spintronics where they can couple to external perturbations ~\cite{gomonay2014spintronics}. In the effective theory for a two sublattice antiferromagnet presented in \cite{dasgupta17}, it was noted that space-time dependent external perturbations introduce gauge fields which can be used to interact with and drive solitons. A similar construction can be envisioned for the three-sublattice case where the solitons in question can be domain walls between the six-fold ground states \cite{yamane2019}.

\begin{acknowledgments}
We are grateful to Collin Broholm, Jonathan Gaudet, and Shu Zhang for illuminating discussions. This work was supported by the U.S. DOE Basic Energy Sciences, Materials Sciences and Engineering Award DE-SC0019331. We acknowledge the hospitality of the Kavli Institute for Theoretical Physics, where this work was supported in part by the  National Science Foundation under Grant No. NSF PHY-1748958. 
\end{acknowledgments}

\appendix
\section{Modifications to elasticity from interlayer interactions}
\label{sec.app.mod.elastic}
The interlayer exchanges are shown in Fig.~\ref{fig:interplane-interaction} and their gradient expanded forms are shown in Eq.~(\ref{eq.U-J}). The interactions expressed using the symmetric vector field $\bm{\beta}$ and the antisymmetric vector field $\bm{\bar\beta}$ contain the following terms :
\begin{enumerate}
\item
A mass term for the field $\bm{\bar\beta}$. 
\item
Direct quadratic interactions: $\partial_i\bm{\bar\beta}\cdot\partial_j\bm{\bar\beta}$ and $\partial_i\bm{\beta}\cdot\partial_j\bm{\beta}$ (`elasticity' theory).
\item
Crossed interaction terms between $\bm{\beta}$ and $\bm{\bar\beta}$ which are linear in derivatives $\bar\beta_i\partial_j\beta_k$. The cross terms have to follow the inversion symmetry criteria for the exchanges.
\end{enumerate}
 Let us take a closer look at the linear term produced by $J_1$ and $J_4$:
\begin{equation}\label{eq.linear_term}
     \mathcal{U}_{\text{linear}} \propto \bar{\beta}_x (\partial_y \beta_x + \partial_x \beta_y) + \bar{\beta}_y (\partial_x \beta_x - \partial_y \beta_y).
\end{equation}
We motivated a generic construction of a six-fold term in Eq.~(\ref{eq:six-fold-anisotropy}). In that construction if we take the vectors $\mathbf{a} = (-\bar\beta_y,\bar\beta_x)$, $\mathbf{b} = \nabla$, and $\mathbf{c} = (\beta_x,\beta_y)$ we generate the cross term in Eq.~(\ref{eq.linear_term}).

\begin{figure}[t]
    \includegraphics[width=\columnwidth]{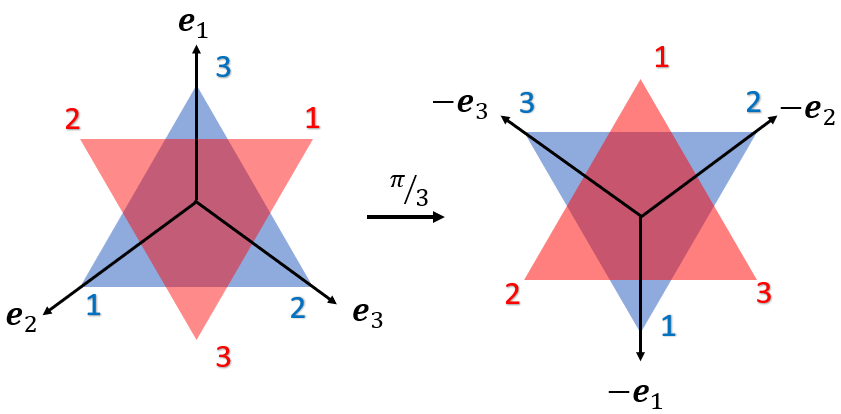}
    \centering
    \caption{This figure shows how a 120$^\circ$ symmetric term converts to a 60$^{\circ}$ term for the central David's star motif in Mn$_3$X. Here we choose the three unit vectors $\mathbf e_1$, $\mathbf e_2$, $\mathbf e_3$ along highly symmetric directions for purposes of illustration. It is clear that after a $\frac{\pi}{3}$ rotation the blue and red (up and down) fields are interchanged and the unit vector axes are reversed. Note that the cyclic permutation of labels caused by the rotation is absorbed into the summation over the labels in Eq.~(\ref{eq:six-fold-anisotropy}).}
    \label{fig:six-fold-sym}
\end{figure}

In section~\ref{sec.three-sl}, we noted that such a term has a $120^{\circ}$ symmetry. For the case of the interlayer coupling this turns into a $60^{\circ}$ symmetry. This happens because in Eq.~(\ref{eq.linear_term}), a $60^{\circ}$ degree rotation interchanges the three unit vectors $\mathbf{e}_i$ with a flipped sign and flips the primed and unprimed fields, which leads to $\bm{\bar\beta} \to -\bm{\bar\beta}$ and $\bm{\beta}\to \bm{\beta}$. The two flips of sign cancel to produce a $60^{\circ}$ symmetry, see Fig.~\ref{fig:six-fold-sym}. 

\begin{figure}[t]
    \includegraphics[width=\columnwidth]{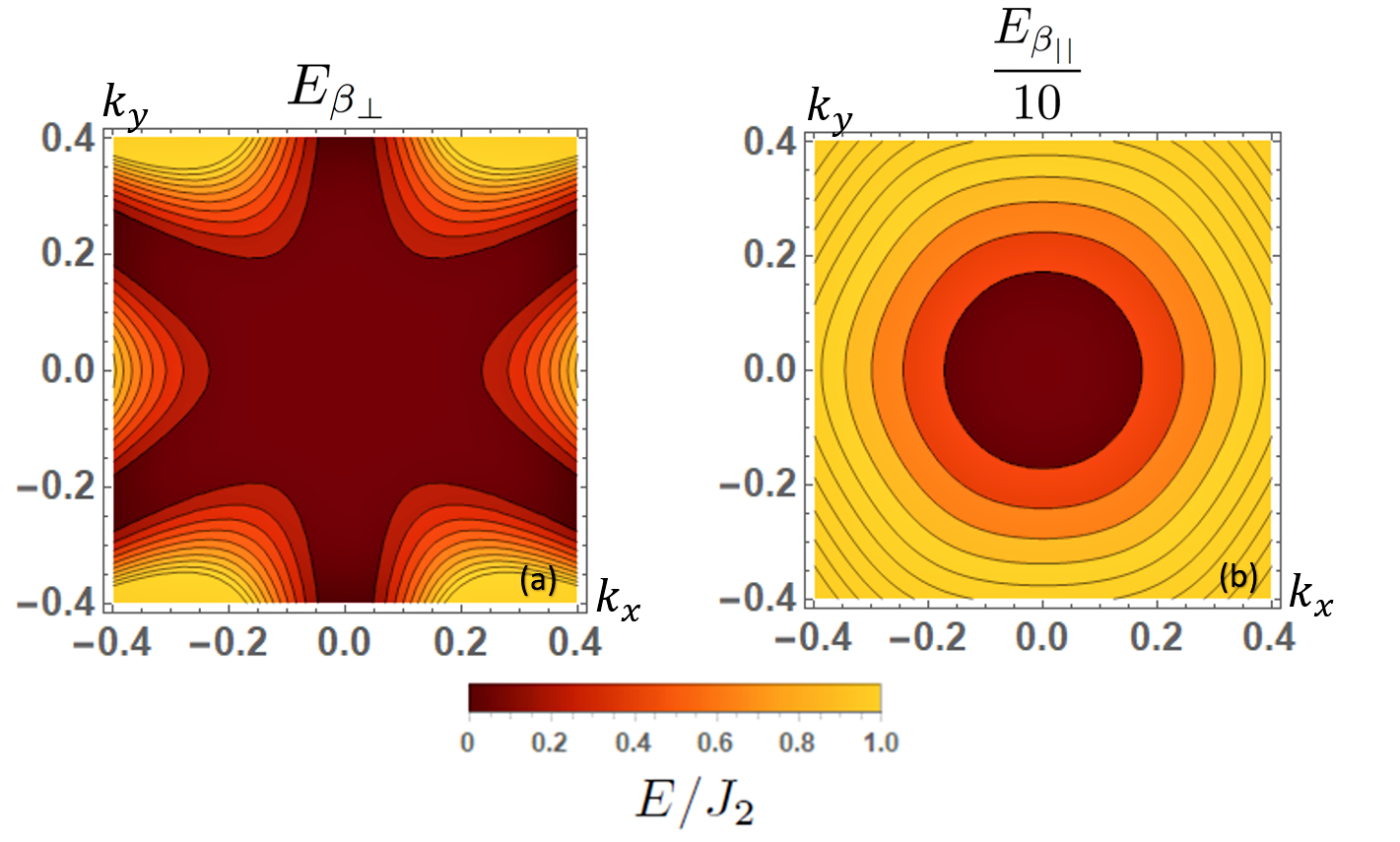}
    \centering
    \caption{Color plots for the calculated dispersions of the $\beta$ doublet with an antiferromagnetic $(J_2,J_1)$ with $J_1 = 2.5~ J_2$, and $J_4 = 0$. (a) : The dispersion for the $\beta_\perp$ mode. (b): The dispersion for the $\beta_{||}$ mode.}
    \label{fig:dispersions}
\end{figure}

This 6-fold symmetry is explicit in the dispersions. Keeping only two antiferromangetic interactions $J_1$ and $J_2$ with $\mathbf{k} = k(\cos\phi_k,\sin\phi_k)$ the two $\bm{\beta}$ modes have the following dispersions to the lowest orders in $k$:
\begin{eqnarray}
    \label{eq.freq-six-fold}
\omega_{\beta_{||}}
&=& Sak \sqrt{(3J_2-J_1)(J_1 + J_2)},
\\
\omega_{\beta_{\perp}} 
&=& \frac{J_1 + J_2}{2} S a^2 k^2 |\cos{3\phi_k}|. \nonumber
\end{eqnarray}
The transverse mode acquires a nonzero frequency, with the exception of six directions, for which $\cos{3\phi_k} = 0$. See Fig.~\ref{fig:dispersions}. In contrast and as apparent in Eq.~(\ref{eq.velocities_total}), $J_4$ has quadratic contributions to both the gapless $\bm{\beta}$ modes  resulting in an isotropic dispersion of the former flat mode. 

\section{Gapping the Goldstones}
\label{sec.app.gaps}
The Goldstone modes are gapped by anisotropies normally present in the kagome magnets Mn$_3$X of these two of them: the easy plane anisotropy, characterized by $\mathcal{K}$, the DM interaction, characterized by the vectors $\mathbf{D}_{ij}$ keep the $U(1)$ symmetry in the $xy$ plane intact. As a result they do not gap the $\alpha_0$ mode and do not split the degeneracy of the $\bm{\beta}$ doublet. The easy plane anisotropy is not included in our model Hamiltonian Eq.~(\ref{eq:H-mnx}), this is done to reduce the number of free parameters in the model. The DM interaction itself provides an easy plane anisotropy which suffices to lift the $\bm{\beta}$ manifold to a finite energy. The local easy-axis anisotropy, characterized by $\delta$, is directed from an Mn site towards the nearest Ge site (at the center of the hexagon) \cite{Chen20}. This interaction breaks the $U(1)$ symmetry of the 120$^\circ$ ground state and gaps the $\alpha_0$ mode and splits the $\bm{\beta}$ doublet. The interactions are given by:
\begin{eqnarray}
\label{eq.energy-anisotropies}
\mathcal{U}_{\text{easy-plane}} &=& \mathcal{K} \sum_{n=1}^3 (\s_n \cdot \mathbf e_z)^2 \\ \nonumber
\mathcal{U}_{\text{DM}} &=&  \sum_{m=1}^3 \sum_{n=1}^3 \mathbf D_{mn} \cdot (\s_m \times \s_n) \\ \nonumber
\mathcal{U}_{\text{easy-axis}} &=& - \delta \sum_{n=1}^3(\s_n \cdot \mathbf e_n)^2
\end{eqnarray}
where the DM vectors $\mathbf D_{mn} = - \mathbf D_{nm} = \pm D \mathbf e_z$ are normal to the easy plane and favor one of the two possible vorticities of spins on a triangle. In both Mn$_3$Sn and Mn$_3$Ge, the antivortex $q=-1$ states are preferred: as we move counterclockwise around a triangle, the spins rotate clockwise.

In the antivortex states, the local anisotropy $U_\text{easy-axis}$ is frustrated: the three magnetization $\s_i$ cannot all point along the respective easy directions. As a compromise, only one of the three sublattices is fully happy, resulting in six possible ground states for each compound. We can express the interactions in Eq.~(\ref{eq.energy-anisotropies}) in terms of the symmetric normal modes $(\alpha_0,\bm{\alpha},\beta_0,\bm{\beta})$. The antisymmetric modes are hardened by a strong $J_4$. 
\begin{widetext}
\begin{eqnarray}
\label{eq.modes-anisotropy}
\mathcal{U}_{\text{easy-plane}} &=& \frac{\mathcal{K} S^2} {2\sqrt{3}a^2 c} \left( \beta_0^2 + \bm{\beta}^2 \right), \\ \nonumber
\mathcal{U}_{\text{DM}} &=& \frac{D S^2}{4 a^2 c} \left( 3\bm{\alpha}^2 + 2\bm{\beta}^2 \right), \\ \nonumber
\mathcal{U}_{\text{easy-axis}} &=& \frac{\delta S^2}{2 a^2 c}\left(\alpha_x\cos{2\phi_0} - \alpha_y\sin{2\phi_0}\right) + \frac{\delta S^2}{4\sqrt{3} a^2 c} \left( \beta_0^2 + \bm{\beta}^2 \right) \\ \nonumber
&-&\frac{\delta S^2}{8\sqrt{3} a^2 c}
\left[
    \left( 
        2\alpha_x^2 - 2\alpha_y^2 + \beta_x^2 - \beta_y^2 + 2\sqrt{2}\beta_0\beta_y 
    \right) 
\cos{2\phi_0} 
    + \left(
        4\alpha_x\alpha_y + 2\beta_x\beta_y + 2\sqrt{2}\beta_0\beta_x 
    \right)
\sin{2\phi_0}  
\nonumber
\right].
\end{eqnarray}
\end{widetext}
Here $\phi_0 = \alpha_0/\sqrt{3}$ is the global rotation angle in the easy plane $ab$. Minimization of the total energy with respect to the three hard modes $\beta_0$ and $\bm{\alpha}$ is again used to eliminate them in favor of the soft modes $\alpha_0$ and $\bm{\beta}$. This procedure yields the energy gaps, Eq. (\ref{eq.alpha_gap}) and Eq.~(\ref{eq.beta_gap}). 

\section{Net spin in the ground state}
\label{sec.app.ferro.mom}
Here we look into a derivation of the Landau functional from which the size of the ferromagnetic moment resulting from spin canting due to $\delta$ can be obtained. Consider a \textit{single kagome layer} with coplanar spins arranged in 120$^\circ$ order in an anticlockwise sense, and an in plane magnetic field. The energy terms we have to consider are: nearest neighbor exchange $J$, easy-axis anisotropy $\delta$, and a Zeeman term.

In each of the six allowed antivortex ground states, the two spins that are not along the local easy-axis try to align along the easy-axis giving rise to a small ferromagnetic moment. This can be expressed in terms of the hard modes $\bm{\alpha}$.
\begin{eqnarray}
\label{eq.fm-mom}
m_x &=& -\sqrt{\frac{3}{2}}S\left(\alpha_x\cos {\phi_0} - \alpha_y\sin {\phi_0}\right), \\ \nonumber
m_y &=& + \sqrt{\frac{3}{2}}S\left(\alpha_x\sin {\phi_0} + \alpha_y\cos{\phi_0}\right).
\end{eqnarray}
Note that in \cite{Chen20} the ground state is at $\alpha_0 \to 0$ in each triangle. Now the size of the moment depends on the values of the doublet $\bm{\alpha}$ in the ground state. To get that we start by writing down the energy density in terms of all six modes:
\begin{eqnarray}
\mathcal{U}_{\text{exchange}} 
&=& \frac{3 J }{2} S^2 \, \bm{\alpha}^2, 
\\ \nonumber
\mathcal{U}_{\text{Zeeman}} 
&=& \sqrt{\frac{3}{2}} \gamma h S
    \left[\alpha_x\cos (\phi_0 + \psi_h) - \alpha_y\sin (\phi_0 + \psi_h)\right], 
\\ \nonumber
\mathcal{U_{\text{easy-axis}}} 
&=& \sqrt{\frac{3}{2}} S^2 \delta 
    \left( \alpha_x\cos{2\phi_0} - \alpha_y\sin{2\phi_0}\right). 
\end{eqnarray}
Here we have used the magnetic field $\mathbf{H} = h (\cos{\psi_h},\sin {\psi_h})$ and $\gamma$ is the gyromagnetic ratio. We can minimize the total energy $\mathcal{U}_{\text{total}} = \mathcal{U}_{\text{exchange}} + \mathcal{U}_{\text{Zeeman}} + \mathcal{U_{\text{easy-axis}}}$ and solve for $\bm{\alpha}$. Plugging the solutions for $\bm{\alpha}$ back into Eq.~(\ref{eq.fm-mom}) we obtain the induced moments as:
\begin{eqnarray}
\label{eq.fm-mom-delta}
m_x &=& \frac{S\delta }{2J} \cos{\phi_0} + \frac{\gamma h }{2J} \cos{\psi_h}, 
\\ \nonumber
m_y &=& \frac{S \delta }{2J} \sin{\phi_0} + \frac{\gamma h }{2J} \sin{\psi_h}.
\end{eqnarray}
Note the extra induced net spin from the anisotropy $\delta$, above the paramagnetic component. For $\phi_0 = 0$ we have $\mathbf{m} = (\frac{S\delta}{2J},0)$ as the ground state in \cite{Chen20} suggests.

\bibliographystyle{apsrev4-1}
\bibliography{main}

\end{document}